\documentclass[emulateapj]{emulateapj}
\newcommand{\ts}{t_{\rm stop}}
\newcommand{\cs}{c_{\rm s}}

\newcommand{\sd}{\Sigma_{\rm d}}

\newcommand{\deldel}[2]{\frac{\partial{#1}}{\partial{#2}} }
\newcommand{\Cw}{C_{\rm w}}
\newcommand{\mdrtot}{{\dot M}_{r,{\rm tot}}}
\newcommand{\tilur}{{\tilde u}_r}
\newcommand{\tilvr}{{\tilde v}_r}
\newcommand{\tvis}{t_{\rm vis}}
\newcommand{\twind}{t_{\rm wind}}
\newcommand{\tdrift}{t_{\rm drift}}
\newcommand{\St}{{\rm St}}

\shorttitle{}
\shortauthors{}

\usepackage{natbib, aas_macros}
\usepackage{amsmath}	% required for `\aligned' (yatex added)
\begin{document}
\title{Structure Formation in a Young Protoplanetary Disk by a Magnetic Disk Wind}

\author{Sanemichi Z. Takahashi\altaffilmark{1,2},Takayuki Muto \altaffilmark{3}}
\altaffiltext{1}{Department of Applied Physics, Kogakuin University,
1-24-2 Nishi-Shinjuku, Shinjuku-ku, Tokyo 163-8677, Japan
sanemichi@cc.kogakuin.ac.jp
}
\altaffiltext{2}{National Astronomical Observatory of Japan, 2-21-1 Osawa, Mitaka, Tokyo 181-8588, Japan.
}
\altaffiltext{3}{
Division of Liberal Arts, Kogakuin University, 1-24-2 Nishi-Shinjuku, Shinjuku-ku, Tokyo 163-8677, Japan
}

\begin{abstract}
Structure formation in young protoplanetary disks is investigated using a one-dimensional model including the formation and the evolution of disks. 
Recent observations with ALMA found that a ring-hole structure may be formed in young protoplanetary disks, even when the disk is embedded in the envelope. 
We present a one-dimensional model for the formation of a protoplanetary disk from a molecular cloud core and its subsequent long-term evolution within a single framework. 
Such long-term evolution has not been explored by numerical simulations due to the limitation of computational power. 
In our model, we calculate the time evolution of the surface density of the gas and the dust with the wind mass loss and the radial drift of the dust in the disk. 
We find that the MHD disk wind is a viable mechanism for the formation of ring-hole structure in young disks. 
We perform a parameter study of our model and derive condition of the formation of ring-hole structures within $6\times 10^5$ years after the start of the collapse of the molecular cloud core. 
The final outcome of the disk shows five types of morphology and this can be understood by comparing the timescale of the viscous diffusion, the mass loss by MHD disk wind and the radial drift of the dust. We discuss the implication of the model for the WL~17 system, which is suspected to be an embedded, yet transitional, disk.
\end{abstract}


\def\bm#1{\mbox{\boldmath $#1$}}

\section{Introduction}
Protoplanetary disks are thought to be the birth place of planets.
The density and temperature structures strongly affect the formation processes of planets.
Recent observations reveal the detailed structures of protoplanetary disk.
They have found that spiral structure 
\cite[e.g.][]{2012ApJ...748L..22M, 2013ApJ...762...48G, 2015A&A...578L...6B, 2016Sci...353.1519P}, non-axisymmetric structure \cite[e.g.][]{2013Sci...340.1199V,2013Natur.493..191C,2013PASJ...65L..14F, 2015PASJ...67..122M}, and ring-like structure \cite[e.g.][]{2007A&A...469L..35G,2010ApJ...725.1735I,2012ApJ...747..136I,2013ApJ...775...30I, 2016PhRvL.117y1101I, 2011ApJ...732...42A,2011ApJ...729L..17H,2012ApJ...758L..19H,2012ApJ...753...59M,2012ApJ...760L..26M, 2015ApJ...808L...3A,2016ApJ...820L..40A, 2016ApJ...829L..35T,2017arXiv171105185F,2017A&A...600A..72F,2017ApJ...840...23L} are formed in protoplanetary disks.

The observed disk structures will give us some clues  to reveal the planet formation scenario.
Especially, the disk structure formed in the early evolutionary phase is important to understand the disk evolution process and the initial condition of the planet formation.
Recently, \cite{2017ApJ...840L..12S} have found that the young protoplanetary disk around WL~17 in $\rho$ Ophiuchus star forming region exhibits a ring structure. 
The dust continuum emission from the disk shows a ring-like structure with a central hole.  
This is similar to the structure of transitional disks.  
However, one remarkable difference between standard transitional disks and the WL~17 system is the disk age.
Observations of \cite{2017ApJ...840L..12S} suggest that WL~17 is still covered by the envelope.
It is regarded as class I YSO \cite[]{2009A&A...498..167V,2009ApJ...692..973E}, whose age is suggested to be $\lesssim 0.5$~Myr \cite[]{2009ApJS..181..321E}, despite large uncertainty.
WL~17 appears like transitional disks, which are one or two orders of magnitude older.
The ring-hole structure formation mechanism has not been well studied for such young disks.
One possibility is the gap formation by (an) unseen planet(s), but it may be very difficult to form planets at such an early stage.
The observations of WL~17 suggest that there may be another mechanism that results in the ring structure formation at the early evolutionary phase.

In this work, we investigate the early phase of disk formation and evolution to explore the possibility of forming small scale structures in young disks.
The important processes for the early evolution of the disk are the gravitational collapse of the cloud core, the angular momentum transfer due to the gravitational instability within the disk, the growth and the radial drift of dust particles, and the effect of the magnetic fields.

The disk formation through the gravitational collapse of the cloud core and the angular momentum transfer due to the  gravitational instability are investigated well by using three-dimensional numerical simulations \cite[e.g.][]{1998ApJ...508L..95B,2007ApJ...670.1198M,2010ApJ...714L..58T,2010ApJ...718L..58I,2011MNRAS.416..591T}.
The subsequent long-term evolution of the star-disk systems have been investigated separately from the star formation phase by assuming some initial disk model. 
This is partly due to the limitation of computational power of solving all the way from the initial star and disk formation from the molecular cloud core to the final dispersal.
However, the evolution of young disks that we focus on in this work may strongly depend on the final outcome of the star-disk formation, which is the initial condition of the disk evolution.

The pioneering theoretical work on the formation and evolution of protoplanetary disks is done by \cite{1981Icar...48..353C} \cite[see also][]{1983Icar...53...26C}.
They have developed the model for the evolution of a one-dimensional viscous accretion disk including the effect of the infall from the cloud core onto the disk.
Following their work, \cite{1994ApJ...431..341S} have done the detailed investigation on the trajectory of accreting gas onto the disk in the early stage of the star-disk formation.
\cite{2005A&A...442..703H} have performed the model calculations with large parameter space of the viscosity, temperature and rotation rate of the cloud cores and compared the results with the observed disk structures.
The model calculations show that the disk becomes gravitationally unstable in their formation phase \cite[]{1994ApJ...421..640N}.
\cite{2010ApJ...713.1143Z} have investigated the disk evolution due to the angular momentum transfer caused by the gravitational instability and MRI using two-layered disk model.
These previous studies used the mass accretion rate from the cloud core onto the disk obtained from the self-similar solution of singular isothermal sphere \cite[]{1977ApJ...214..488S}, which gives time-independent mass accretion rate. 
In \cite{2013ApJ...770...71T}, we have constructed the analytical model providing the time-dependent mass accretion rate from the cloud core with arbitrary radial density profile.

The importance of dust growth and radial drift at the initial disk formation stage has been pointed out recently.
Two-dimensional numerical simulations show that dust particles grow and drift inward resulting in the small dust-to-gas mass ratio during the disk formation \cite[]{2018arXiv180106898V}.
The dust growth and the decrease of dust-to-gas ratio in the disk formation stage are also indicated by the steady accretion disk model \cite[]{2017ApJ...838..151T}.
Such behavior of dust particles in the disk formation phase will affect the following disk evolution. 
To investigate the early evolution of disks, we need to deal with the gravitational collapse of the cloud core, the disk evolution caused by the angular momentum transfer, and the dust evolution in the disk comprehensively.

In addition, we investigate the effects of the disk wind that can be driven by the magneto-rotational instability \cite[]{2009AIPC.1158..161S, 2010ApJ...718.1289S} on the formation of small scale structures in young disks.
\cite{2010ApJ...718.1289S} investigated the mass loss rate due to the disk wind. They parameterized the mass loss by $\Cw$, where ${\dot \Sigma}_{\rm wind}=\Cw \Sigma \Omega$.
According to the three-dimensional local MHD simulations, $\Cw\sim 10^{-5}-10^{-3}$.
The timescale of wind mass loss scales with the local Kepler time of the disk, which is faster at inner radii.
Therefore, if $\Cw$ is constant in the disk, 
the wind mass loss is efficient in the inner region and the inner hole is naturally formed by the wind.
In such disks, it is expected that the dust concentrates around the inner edge of the disk and the ring-hole structure in dust distribution will be formed.
The long-term evolution of the disks including the disk wind is investigated by using one-dimensional models treating isolated disks that have already been formed \cite[e.g.][]{2016A&A...596A..74S,2016A&A...596A..81P}.
These work, however, did not calculate the disk formation phase.

In this paper, we calculate the formation and the evolution of protoplanetary disks within a single framework.
We extend the model provided in \cite{2013ApJ...770...71T}, which takes into account the time-dependent mass accretion, to include the effect of the dust and the wind mass loss on the disk evolution. 
We show that  various gas and dust distributions can be formed by the MRI disk wind in young protopranetary disks.

This paper is organized as follows. The model for the formation and the evolution of the disk is explained in Section \ref{method}. 
In Section \ref{result}, we show the result obtained from the model.
Section \ref{discussion} and \ref{conclusion} are discussion and conclusion.

\section{Method}
\label{method}
In this work, we calculate the formation and the evolution of protoplanetary disks to investigate the ring structure formation in young disks.
Since we focus on young disks whose ages are similar to the disk formation timescale ($\sim 10^6$ yr), we cannot assume the already formed protoplanetary disks, for example minimum mass solar nebulae, as initial stage.
Instead, we adopt molecular cloud cores as initial conditions and calculate the gravitational collapse of the cloud core and the disk evolution simultaneously.

\begin{figure}[tb]
 \includegraphics[width=9cm]{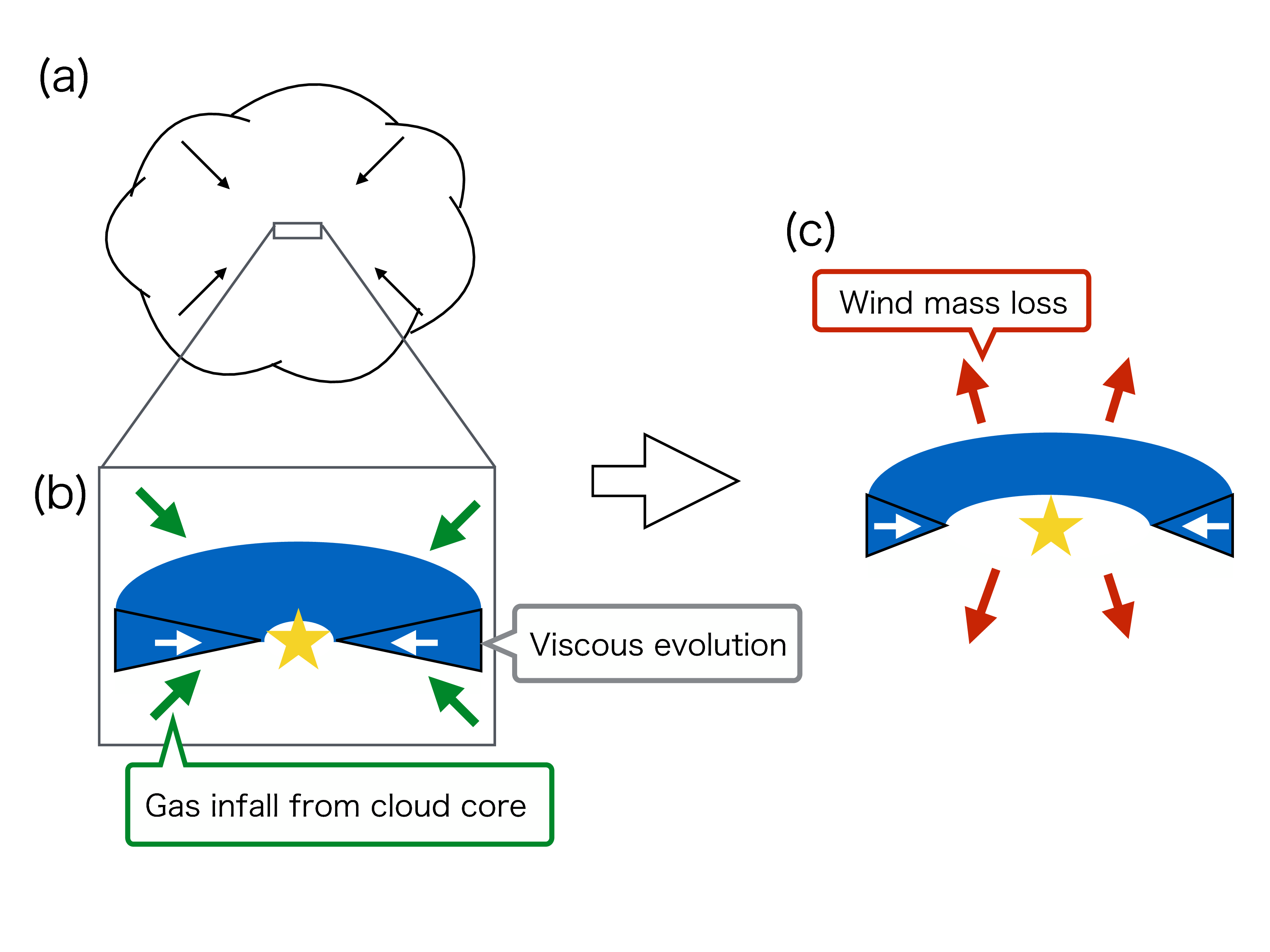}
\caption{Schematic picture of the formation and evolution of the protoplanetary disk that we calculate in this work.}
\label{fig:schematic}
\end{figure}

Figure \ref{fig:schematic} shows the schematic picture of the formation and evolution of protoplanetary disks.
We calculate the gravitational collapse of the cloud core by a one-dimensional semi-analytic model (Panel (a) of Figure \ref{fig:schematic}).
The gas around the center of the core has small angular momentum so that it makes the protostar through the gravitational collapse.
The gas in the outer region of the cloud core has large angular momentum.
Thus, it cannot directly falls onto the protostar but makes the disk around the protostar.
Here, we assume that the gas falls at the centrifugal radius of the disk.
Using the semi-analytic model, we obtain the mass infall rate per unit area on the protoplanetary disk \cite[]{2013ApJ...770...71T}.
We also calculate the evolution of the disk with the mass infall from the cloud core (Panel (b) of Figure \ref{fig:schematic}).
After the end of the collapse of the cloud core, we take into account the wind mass loss from the disk (Panel (c) of Figure \ref{fig:schematic}).
In this section, we show the equations to calculate the gravitational collapse of the cloud core (Panel (a) of Figure \ref{fig:schematic}) and the evolution of the protoplanetary disks (Panel (b) and (c) of Figure \ref{fig:schematic}).

 \subsection{Formulation}
\subsubsection{Gravitational collapse of cloud cores and gas infall onto disks}
First, we describe the one-dimensional semi-analytic model to calculate gravitational collapse of cloud cores (Panel (a) of Figure \ref{fig:schematic}). 
Here, we use the models given in \cite{2013ApJ...770...71T} (see also \cite{2016MNRAS.463.1390T} for a detailed derivation).
The validity of the model is tested by the comparison with a three-dimensional numerical simulation in \cite{2013ApJ...770...71T, 2016MNRAS.463.1390T} and we found that the model reproduces the results of the numerical simulation well.
In this model, we first evaluate the total mass accretion rate onto the central star and the disk to obtain the increase rate of the disk surface density per unit time $\dot{\Sigma}_{\rm inf}$ due to the mass infall from the cloud core.
Since the typical scale of the cloud cores ($\sim 0.1$ pc) is much larger than that of the disk ($\sim 100 $ au), we neglect the centrifugal force and assume the spherical symmetry in the modeling of the collapsing cloud core.
We adopt the mass accretion rate onto the center of the cloud core as the total accretion rate onto the star and the disk.
We assume that molecular cloud cores and infalling envelopes are isothermal (10 K).
The initial density profile is given so that the initial gravitational force is larger than the initial pressure gradient force by a factor of $f$.
Because of the pressure gradient force, the time when the collapsing gas reaches the disk $t_{\rm inf}$ is larger than the free-fall time $t_{\rm ff}$ (Equation (6) in \cite{2016MNRAS.463.1390T}):
\begin{equation}
 t_{\rm inf}=\frac{2}{\pi}t_{\rm ff}\int^1_0
  \frac{dx}{\sqrt{f^{-1}\ln x+ x^{-1}-1}}.
\label{eq:tinf}
\end{equation}
Since the free-fall time $t_{\rm ff}$ depends on the initial radius of the cloud core, the infall time $t_{\rm inf}$ also depends on the initial radius of the gas in the core.
The mass accretion rate onto the disk is given by 
\begin{equation}
{\dot M}_{\rm inf}(t) = 4\pi R_{\rm ini}^2\rho_{\rm ini}(R_{\rm ini})\left(\left.\frac{dt_{\rm inf}}{dR_{\rm ini}}\right|_{t_{\rm inf}=t}\right)^{-1},
\label{eq:Mdotinf}
\end{equation}
where $R_{\rm ini}$ is the initial radius of the infalling gas and $\rho_{\rm ini}(R_{\rm ini})$ is the initial density of the cloud core at radius $R_{\rm ini}$.

We assume that cloud cores initially rotate rigidly and the specific angular momentum of the infalling gas is conserved.
To derive the increase of the surface density of the disk per unit time ${\dot \Sigma}_{\rm inf}$, we assume the envelope gas falls at the centrifugal radius \cite[see also Equation (10) in][]{2013ApJ...770...71T}:
\begin{equation}
 {\dot \Sigma}_{\rm inf} = \frac{1}{2\pi r}\frac{{\dot M}_{\rm inf}}{2\Omega_{\rm core} R_{\rm ini}^2}\left(1-\frac{j}{\Omega_{\rm core} R_{\rm ini}^2}\right)^{-1/2} \deldel{j}{r},
\label{eq:sigma_inf}
\end{equation}
where $\Omega_{\rm core}$ is the initial angular velocity of the cloud core, $r$ is the disk radius, and $j$ is the specific angular momentum distribution of the disk, which is related with the disk angular velocity $\Omega$ by $j=r^2\Omega$.
This equation is the same as Equation (6) in \cite{2005A&A...442..703H} if $j\propto r^{1/2}$.
In this work, we assume that the total molecular cloud core mass $M_{\rm core} = 1.5M_\odot$. We stop the infall when the all of gas in the cloud core falls onto the central star or the disk.

\subsubsection{Disk evolution}
Due to the gas infall from the core, protoplanetary disks are formed.
In order to investigate the small structure formation in young disks, we calculate the time evolution of the surface densities of the gas and the dust (Panel (b) and (c) of Figure \ref{fig:schematic}).
We assume that the disk evolution timescale is much larger than the Kepler timescale.
The radial motion of dust due to the interaction with the gas has been well studied \cite[cf.][and the references therein]{1972fpp..conf..211W,1977MNRAS.180...57W, 2016SSRv..205...41B}.  The one-dimensional model of gas and dust evolution is obtained for both inviscid and viscous disks \cite[cf.][]{1986Icar...67..375N, 2009ApJ...690..407K, 2017ApJ...844..142K}. 
In this work, we extend the model to the disk with the mass accretion from the cloud core and the wind mass loss.
The equations for the surface densities and the radial velocities of the gas and the dust are given as follows (a derivation of the equations are given in Appendix),
\begin{equation}
 \deldel{\Sigma}{t}=-\frac{1}{r}\deldel{}{r}r\Sigma u_r
  +{\dot \Sigma}_{\rm inf} 
  -{\dot \Sigma}_{\rm wind},
\label{eq:dSigma_dt}
\end{equation}
\begin{equation}
 \deldel{\sd}{t} = -\frac{1}{r}\deldel{}{r}r\sd v_r + \epsilon{\dot \Sigma}_{\rm inf},
\end{equation}
\begin{eqnarray}
 u_r
&=&\frac{2}{r\Omega}N
-\frac{\sd}{\Sigma+\sd}\frac{1}{A\St'^2+1}
\frac{2}{r\Omega}N \nonumber \\
&&+\frac{2\sd}{\Sigma+\sd}\frac{A\St'}{A\St'^2+1}\eta r \Omega - \frac{r\mdrtot}{M_r}
\label{eq:ur}
\end{eqnarray}
\begin{eqnarray}
 v_r & =& \frac{\Sigma}{\Sigma+\sd}\frac{1}{A\St'^2+1}
\frac{2}{Ar\Omega}N
-\frac{\Sigma}{\Sigma+\sd}\frac{2\St'}{A\St'^2+1}\eta r \Omega \nonumber \\
&&+\left(\frac{2}{Ar\Omega}N-\frac{\sd}{\Sigma+\sd}\frac{1}{A\St'^2+1}\frac{2}{Ar\Omega}N \right.\nonumber \\
&&\left. \ \ \ \ \ \ +\frac{2\sd}{\Sigma+\sd}\frac{\St'}{A\St'^2+1}\eta r \Omega \right)\frac{2\pi r^2 \Sigma}{M_r} - \frac{r\mdrtot}{M_r}
\label{eq:vr}
\end{eqnarray}
where $\Sigma$ and $u_r$ are the surface density and the radial velocity of the gas, $\sd$ and $v_r$ are those of the dust, ${\dot \Sigma}_{\rm wind}$ is the mass loss rate per unit area due to the disk wind, $\epsilon$ is the dust-to-gas mass ratio in the infalling envelope, $\Omega$ is the angular velocity of gas and dust, $M_r$ is the enclosed mass within the radius $r$.
The specific torque $N$ is given by
\begin{equation}
 N=\frac{1}{r\Sigma}\deldel{}{r}\left(
r^3\nu\Sigma\deldel{\Omega}{r}
\right),
\label{eq:T}
\end{equation}
where $\nu$ is the coefficient of the kinematic viscosity.
Here, we assume that the viscosity is caused by the gravitational instability and magnetorotational instability (see Equation (\ref{eq:alpha})).
The factor $A$ represents the effect of the self-gravity of the disk of gas,
\begin{equation}
 A=\left(1+\frac{2\pi r^2\Sigma}{M_r}\right).
\label{eq:A}
\end{equation}
The Stokes number St is 
\begin{equation}
 \St=\sqrt{\frac{\pi}{8}}\frac{\rho_{\rm i} a \Omega}{\rho \cs} = \frac{\pi \rho_{\rm i}a}{2\Sigma},
\label{eq:St}
\end{equation}
where $a$ is the radius of the dust, $\rho_{\rm i}$ is the internal density of the dust and $\rho$ is the gas density at the midplane of the disk. We assume $\rho_{\rm i}=3\ {\rm g\ cm^{-3}}$.
$\St'$ is the modified Stokes number,
\begin{equation}
 \St'=\frac{\Sigma}{\Sigma+\sd}\St.
\label{eq:St'}
\end{equation}
We evaluate $\eta$ as follows,
\begin{equation}
 \eta=-\frac{1}{2}\left(\frac{\cs}{r\Omega}\right)^2
\deldel{\ln p}{\ln r}.
\label{eq:eta}
\end{equation}
where $p$ is the pressure at the disk midplane and $\cs$ is the isothermal sound speed with mean molecular weight $\mu=2.34$.
The mass input/loss rate due to the infall and disk wind within the radius $r$ $\mdrtot$ is 
\begin{equation}
 \mdrtot = \int^r_02\pi r
({\dot \Sigma}_{\rm inf}-{\dot \Sigma}_{\rm wind})dr.
\label{eq:mrdot}
\end{equation}

We assume the centrifugal balance in the disk and that the gravitational force is approximately given by $-GM_r/r^2$ and the pressure gradient force, the frictional force and the viscosity are small compered with the centrifugal force and the gravitational force.
Thus, the angular velocity distribution in the disk is given by
\begin{equation}
 \Omega=\sqrt{\frac{G M_r}{r^3} }.
\label{eq:centrifugal_balance}
\end{equation}
As shown in Equation (\ref{eq:T}), we evaluate the torque by using the effective viscosity represented by $\nu = \alpha\cs^2/\Omega$, where $\alpha$ is a dimensionless measure of turbulent intensity \cite[]{1973A&A....24..337S}. 
We assume the locally isothermal in the envelope and the disk, which are given in Section \ref{modelsetup}.

In this work, we take into account the angular momentum transport caused by gravitational instability and  magnetorotational instability \cite[MRI,][]{1991ApJ...376..214B},
\begin{equation}
 \alpha = \alpha_{\rm GI}+\alpha_{\rm MRI}.
\label{eq:alpha}
\end{equation}
The $\alpha_{\rm GI}$ is given by a function of the Toomre's $Q$ parameter 
\begin{equation}
 \alpha_{\rm GI} = \exp(-Q^4),
\label{eq:alphaGI}
\end{equation}
\begin{equation}
 Q=\frac{\cs\Omega}{\pi G \Sigma}.
\label{eq:Q}
\end{equation}
This effective viscosity becomes efficient when the disk becomes gravitationally unstable \cite[]{2010ApJ...713.1143Z}.
This model can mimic the disk evolution due to the angular momentum transfer by spiral arms formed in gravitationally unstable disks \cite[]{2013ApJ...770...71T}.
We assume that $\alpha_{\rm MRI}$ is constant in the disk.

We use the simple model to evaluate the mass loss rate due to the disk wind ${\dot \Sigma}_{\rm wind}$.
We assume that the disk wind becomes efficient after the infall is finished. 
We discuss the validity of this treatment in Section \ref{windmodel}.
We introduce the efficiency parameter $\Cw$ for the wind mass loss as follows
\begin{equation}
 {\dot \Sigma}_{\rm wind} = 
\begin{cases}
 0 & M_{\rm inf}<M_{\rm core}, \\
\Cw\Sigma\Omega & M_{\rm inf}=M_{\rm core},
\end{cases}
\end{equation}
where $M_{\rm inf}$ is the mass that has already fallen onto the central star and the disk.
Since the gas density decreases with increasing height and that at the launching point of the wind is a factor of $\sim 100$ smaller than that on the midplane, the coupling between the gas and the dust in the wind is much weaker than that on the midplane. 
The dust grains are blown out by the wind only when they are small enough to be coupled with the gas in the wind.
\cite{2016ApJ...821....3M} have shown that the dust grains with $St \lesssim 10^{-6}$ can be blown out by the disk wind.
Since such grains are much smaller than that we adopt in this work (see Table \ref{tab:param}), the dust mass loss rate is assumed to be zero in our model.
\footnote{The small dust blown out by the wind will be the origin of the continuum emission observed in the outflow\cite[cf.][]{2003A&A...401L...5G}. 
When we calculate the time evolution of the dust size distribution and wind mass loss of the dust, we can compare the model with the observations and test the validity of the model of the dust wind mass loss.
}

The list of symbols used for equations in this paper is shown in Table \ref{tab:symbols}.
\begin{table}
\caption{List of symbols. Last 10 symbols are parameters in our model.}
\label{tab:symbols}
\scalebox{1}[1]{
\begin{tabular}{ll}
\hline
$t_{\rm inf}$ & time that the envelope reaches center, Equation (\ref{eq:tinf}) \\
$t_{\rm ff}$ & free fall time of the envelope\\
${\dot M}_{\rm inf}$ & mass infall rate into the disk, Equation (\ref{eq:Mdotinf})\\
$R_{\rm ini}$ & initial radius of the infalling gas \\
$\rho_{\rm ini}$ & initial density of the infalling gas \\
$j$ & specific angular momentum \\
$\Omega$ & angular velocity \\
$\Sigma$ & surface density of gas \\
$u_r$ & radial velocity of gas \\
$\sd$ & surface density of dust \\
$v_r$ & radial velocity of dust \\
${\dot \Sigma}_{\rm inf} $ & mass infall rate per unit area \\
${\dot \Sigma}_{\rm wind}$ & wind mass loss rate per unit area\\
$\nu$ & coefficient of the kinematic viscosity\\
$M_r$ & enclosed mass within the radius $r$\\
$T$ & specific torque, Equation (\ref{eq:T})\\
$A$ & Equation (\ref{eq:A})\\
$a$ & dust radius\\
$\rho_{\rm i}$ & internal density of dust \\
$\rho$ & gas density at the midplane of the disk \\
St' & modified Stokes number, Equation (\ref{eq:St'})\\
$\eta$ & Equation (\ref{eq:eta})\\
$p$ & pressure at the disk midplane \\
$\cs$ & isothermal sound speed \\
$\mdrtot$ & mass infall/loss rate within $r$ \\
$\alpha$ & Equation (\ref{eq:alpha}) \\
$\alpha_{\rm GI}$ & Equation (\ref{eq:alphaGI})\\
$Q$ & Toomre's parameter, Equation (\ref{eq:Q})\\
$M_{\rm inf}$ & total infalled mass \\
$T_{\rm eq}$ & temperature of the disk, Equation (\ref{eq:Teq})\\
\hline
\hline
$f$ & mass enhancement factor of the core\\
$\Omega_{\rm core}$ & initial angular velocity of the core\\
$M_{\rm core}$ & total mass of the cloud core\\
$\epsilon$ & dust-to-gas mass ratio in the envelope\\
St & Stokes number, Equation (\ref{eq:St})\\
$\mu$ & mean molecular weight\\
$\alpha_{\rm MRI}$ & dimensionless measure of MRI turbulent intensity\\
$\Cw$ & efficiency parameter for wind mass loss\\
$\rho_0$&initial central density of the core  \\
$T_{\rm core}$&temperature of the core\\
\hline
\end{tabular} 
}
\end{table}

\subsection{Boundary Conditions}
In the calculations, we set the outer boundary at $10^4$ au and the inner boundary at 1 au.
At the outer boundary, we assume $u_r=v_r=0$.
For the inner boundary, we assume $\partial \Omega/\partial r =0$ at the center to evaluate $N$.
We also assume $\partial \ln p /\partial \ln r$ is constant at the inner boundary.

\subsection{Model setup}
\label{modelsetup}
Since the calculations start from gravitational collapse of cloud cores, we adopt the density and rotation profiles of cloud cores as initial conditions.
The density distributions of cloud cores are given by the Bonner-Evert sphere with the central density is $\rho_0=10^{-18}\ {\rm g\ cm^{-3}}$ and the temperature is $T_{\rm core}=10$ K.
In order to make cores gravitationally unstable, we increase the density of the BE sphere by a factor of $f=$1.4.
The rotation velocity of the cores is given by the rigid rotation with the angular velocity $\Omega_{\rm core}=$ 0.3 ${\rm km \ s^{-1}\  pc^{-1}}$ (the dependence of the disk evolution on $\Omega_{\rm core}$ is discussed in Section \ref{jcore}).

We assume that the temperature is given by the equilibrium temperature $T_{\rm eq}$ that is obtained from the balance between the irradiation heating from the central star and the radiation cooling at the disk surface \cite[]{1997ApJ...490..368C},
\begin{equation}
 T_{\rm eq} = {\rm max}
\left[ 150\left(\frac{r}{1\ [{\rm au}]}\right)^{-3/7}
, 10\right] \ [{\rm K}].
\label{eq:Teq}
\end{equation}
The parameters used in this work are shown in Table \ref{tab:param}.

Due to dust growth in the disk, the typical dust size depends on the disk radius.
In the inner region, the small dust grows quickly. 
The resulting dust size distribution is roughly given by the constant Stokes number of $\sim 0.1$ \cite[]{2012ApJ...752..106O}.
We therefore assume that the Stokes number is the same at all radii of the disk and  set the fiducial value of the Stokes number to be 0.1, but we explore the parameter space of St.
The disk evolution in the case of the constant dust radius is discussed in Section \ref{a_dust}.
The smallest value of St=$10^{-3}$ corresponds to the tight coupling between gas and dust resulting in almost constant dust-to-gas mass ratio in the disk. 
We set the largest Stokes number to be 10 times larger than the fiducial value: St=1.
The largest value corresponds to the size for which the interaction between the gas and the dust is the most efficient.
Note that the fiducial value of the Stokes number corresponds to the $\sim$ mm size grains in the ring structure, which is the size probed by mm- and submm- wavelengths observations (see Section \ref{WL17}).
Since the strength of the turbulence and the wind mass loss is quite uncertain, we also explore the parameter space of $\alpha_{\rm MRI}$ and $\Cw$ obtained in the numerical simulations on the MRI and disk wind \cite[e.g.][]{2010ApJ...718.1289S,2011ApJ...742...65O}.
The fiducial parameters that we involve in this work is summarized in Table \ref{tab:param}.

\begin{table}
\caption{List of parameters}
\label{tab:param}
\begin{center}
\begin{tabular}{lll}
\hline
&Fiducial value& Range\\
\hline
$\alpha_{\rm MRI}$ & $3\times10^{-4}$ & $10^{-5}-10^{-2}$\\
$\Cw$ &  $10^{-4}$ & $10^{-5}-10^{-2}$\\
St & 0.1 &$10^{-3}-1$\\
\hline
$f$ & 1.4& \\
$\Omega_{\rm core}$ &  0.3 [${\rm km\ s^{-1}\ pc^{-1}}$]& \\
$M_{\rm core}$ & 1.5$M_{\rm \odot}$& \\
$\epsilon$ & 0.01 &fixed\\
$\mu$ & 2.34 &\\
$\rho_0$ & $10^{-18}\ [{\rm g\ cm^{-3}}]$ &\\
$T_{\rm core}$ & 10 [K] &\\
\hline
\end{tabular} 
\end{center}
\end{table}

\section{Results}
\label{result}

\subsection{Disk Evolution and Ring Formation}
First of all, we present the result of a model with one set of parameters, where we observe the formation of a ring structure. 
The parameters are $\alpha_{\rm MRI} = 3\times 10^{-4},\ \Cw=10^{-4}$ and St=0.1.

\begin{figure}[tb]
 \includegraphics[width=9cm]{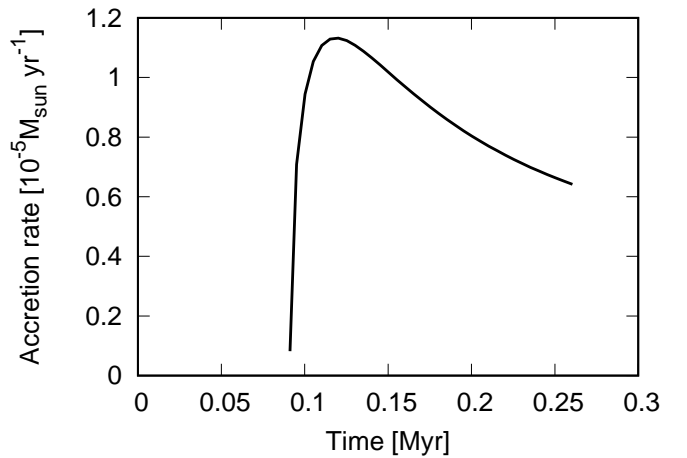}
\caption{The mass infall rate onto the disk and the central star from the cloud core. The infall rate does not depend on the parameters  $\alpha_{\rm MRI},\ \Cw$ and St.}
\label{fig:Macc}
\end{figure}

\begin{figure}[tb]
 \includegraphics[width=9cm]{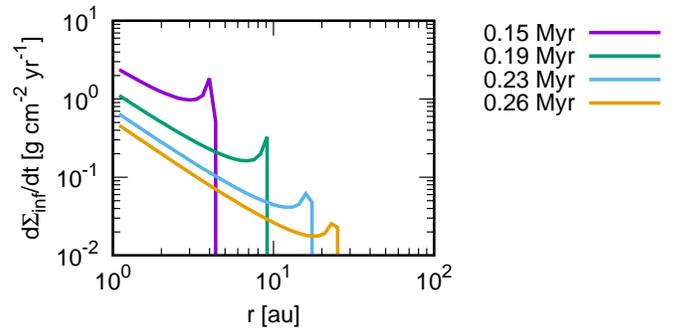}
\caption{Time evolution of the mass infall rate per unit area.}
\label{fig:Macc_prof}
\end{figure}

\begin{figure}[tb]
\begin{tabular}{c}
\includegraphics[width=9cm]{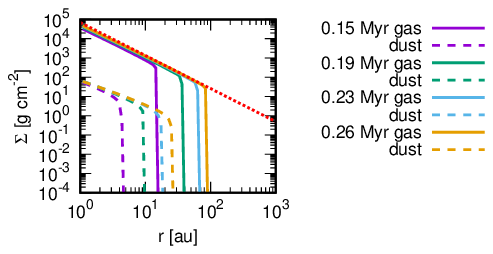}
 \\
\includegraphics[width=9cm]{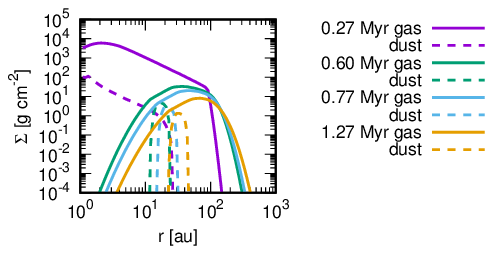}
\end{tabular}

\caption{Time evolution of the surface density of the gas and dust with $\alpha_{\rm MRI} = 3\times 10^{-4},\ \Cw=10^{-4}$ and St=0.1. The solid lines shows the gas surface density and the dotted lines shows the dust surface density. The top panel shows the surface densities before the infall stops and the bottom panel shows those after the infall stops.
The red dotted line in the upper panel shows the surface density satisfying $Q=1$ given in Equation (\ref{eq:sigma_Q1}).}
\label{fig:sigma_st1}
\end{figure}

We first look at the mass infall phase where a protoplanetary disk is formed around a forming central star. 
Figure \ref{fig:Macc} shows the mass infall rate from the core onto the disk and the central star.
The infall rate does not depend on the parameters  $\alpha_{\rm MRI},\ \Cw$ and St but depends only on the density and temperature profiles of the cloud core.
The infalling gas reaches the center of the cloud core at $t\sim0.1$~Myr, which is the time of the protostar formation.
The gas infall continues until $\sim 0.27$~Myr.
The time evolution of the mass infall rate per unit area is shown in Figure \ref{fig:Macc_prof}. 
In this case, the angular momentum of the cloud core is small and all the gas infalls within $\sim 30 $ au.
The size of the disk depends on the angular velocity of the core, which will be discussed in Section \ref{jcore}.
The top panel of Figure \ref{fig:sigma_st1} shows the time evolution of the surface density of the gas and dust before the gas infall from the cloud is finished at $t<0.27$~Myr.
The disk is kept gravitationally unstable during this phase since gas is continuously supplied from the envelope. 
Thus, the angular momentum is redistributed mainly due to the effect of the gravitational instability (Equation (\ref{eq:alphaGI})).
The gas accretes inward and/or expands outward resulting in the decrease of the surface density  while the infall from the cloud increases it.
As a result, the disk with infall from cloud core sustains $Q\sim1$ \cite[cf.][]{2004MNRAS.351..630L,2007MNRAS.381.1009V,2015MNRAS.446.1175T,2016MNRAS.458.3597T}.
Using the central star mass $1M_{\odot}$ and temperature distribution given in Equation (\ref{eq:Teq}), the radial distribution of the surface density with $Q=1$ is 
\begin{equation}
\Sigma=6.9\times 10^4 \left(\frac{r}{1 [\rm au]}\right)^{-12/7}
\ [{\rm g\ cm^{-2}}].
\label{eq:sigma_Q1}
\end{equation}
This surface density distribution is shown as a red dotted line in the top panel of Figure \ref{fig:sigma_st1}.
Note that the gas disk radius is larger than the maximum centrifugal  radius of the infalling gas.
The radius of the disk expands due to the outward mass flux caused by the angular momentum transfer in the disk.
On the other hand, dust disk radius is almost same as the maximum centrifugal radius of the infalling gas.
This is because the dust decouples from the gas and does not expand outward due to the large Stokes number St=0.1.
Moreover, the radial drift of the dust causes the smaller dust-to-gas mass ratio than that of the infalling envelope $\epsilon=0.01$.

\begin{figure}[tb]
\includegraphics[width=9cm]{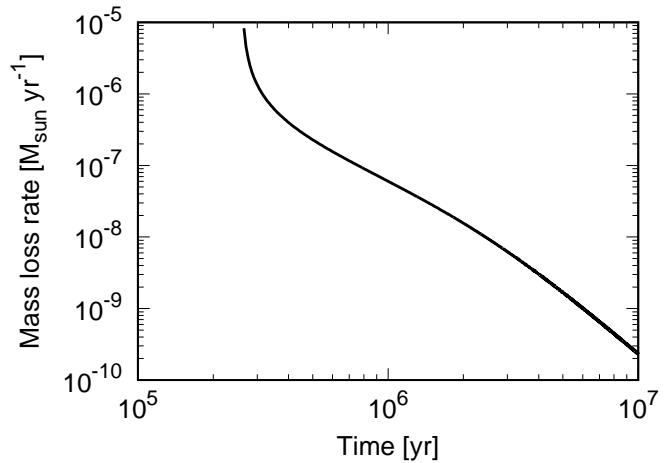}
\caption{Mass loss rate obtained from our model calculation with  $\alpha_{\rm MRI}=3\times10^{-4}$, $\Cw=10^{-4}$, St=0.1.}
\label{fig:Mdotwind}
\end{figure}

We now turn our attention after the mass infall from the cloud core is finished at $t=0.27$~Myr.  We note that our model calculations continue from the infall phase, although we present the results before and after the infall stops separately. 
The bottom panel of Figure \ref{fig:sigma_st1} shows the evolution of the surface density of the gas and the dust.
The time $t=0.27$~Myr is just after the infall stops and the wind mass loss stars.
In Figure \ref{fig:Mdotwind}, we show the time evolution of the mass loss rate.
The wind mass loss make the ring-hole structure in the gas disk. 
The dust accumulates at the radius of the pressure maximum around the inner edge of the gas ring structure.
The dust ring structure at the radius of $\sim 10$ au with the width of $\sim 10$ au is formed at $t=0.6$~Myr, which corresponds to the protostar age (calculated from 0.1~Myr) of $\sim 0.5$~Myr.

\begin{figure}[tb]
\includegraphics[width=9cm]{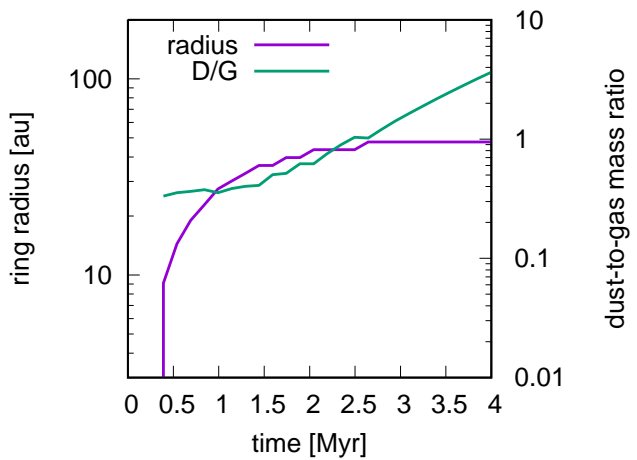}
\caption{Ring radius and dust-to-gas mass ratio at the ring center with $\alpha_{\rm MRI}=3\times10^{-4}$, $\Cw=10^{-4}$ and St=0.1. The purple line shows the ring radius and the green line shows the dust-to-gas mass ratio at the radius of the maximum dust surface density.}
\label{fig:ring_timeevo1}
\end{figure}

Figure \ref{fig:ring_timeevo1} shows the ring radius and dust-to-gas mass ratio at the center of the ring with $\alpha_{\rm MRI}=3\times10^{-4}$, $\Cw=10^{-4}$ and St=0.1.
The ring radius increases with time because the radius of the pressure maximum increases due to the wind mass loss of gas.
The dust-to-gas mass ratio is about 0.3 for $t\lesssim 1.5$~Myr.
For $t>1.5$~Myr, the dust-to-gas mass ratio increases
because 
significant amount of gas is depleted due to the disk wind while the dust particles remain in the disk.
The expansion of the ring slows down as dust particles condensate.
To move the dust outward in the disk, the gas need to give the angular momentum to the dust.
However, when the dust-to-gas mass ratio increases, the gas has no enough inertia to move the dust.
Thus, the ring radius does not increase after the time when the dust-to-gas mass ratio become larger than unity.
This behavior appears only when we use the equations taking into account the back reaction  from dust to gas.

\subsection{Parameter Study}
\begin{figure*}[tb]
\begin{center}
\includegraphics[width=15cm]{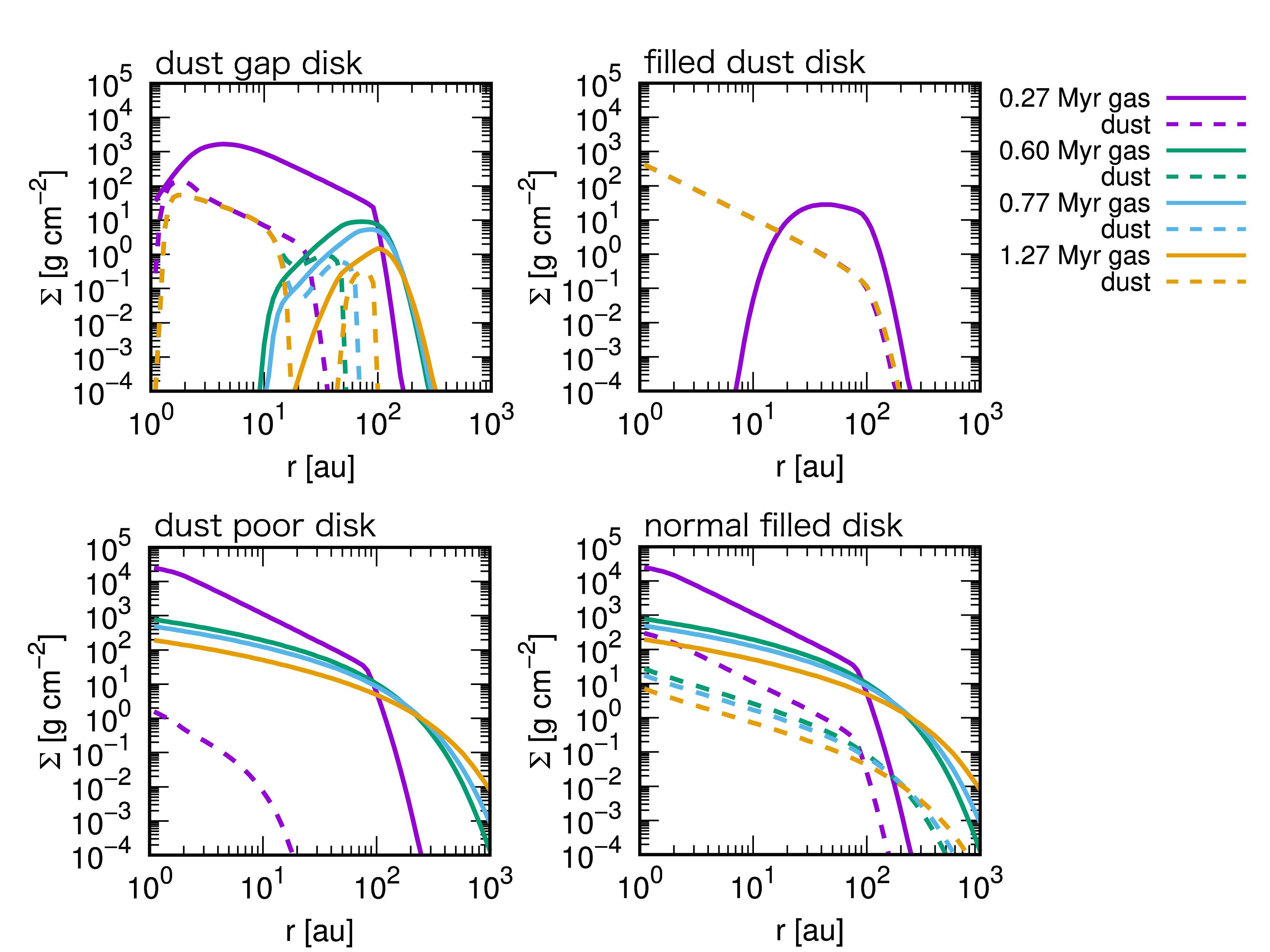}
\caption{Time evolution of the surface density of gas and dust with $\alpha_{\rm MRI}=10^{-4}$, $\Cw=3\times10^{-4}$ , St=0.03 (top left, dust gap disk),
$\alpha_{\rm MRI}=10^{-5}$, $\Cw=10^{-2}$, St=$10^{-3}$ (top right, filled dust disk),
$\alpha_{\rm MRI}=10^{-2}$, $\Cw=10^{-5}$, St=1 (bottom left, dust poor disk), 
$\alpha_{\rm MRI}=10^{-2}$, $\Cw=10^{-5}$, St=$10^{-3}$ (bottom right, normal filled disk).
}
\label{fig:sigma_type}
\end{center}
\end{figure*}

The gas and dust structures formed in disks depend on the parameters $\alpha_{\rm MRI}$, $\Cw$, and St.
In this section, we show other disk structures formed in the disk.
Figure \ref{fig:sigma_type} shows four typical outcomes of the disk structures obtained form our model calculations: the dust gap disk, the filled dust disk, the dust poor disk, and the normal filled disk are formed.

In the case of the dust gap disk ($\alpha_{\rm MRI}=10^{-4}$, $\Cw=3\times10^{-4}$, St=0.03), the dust ring is formed at $r\gtrsim 20$ au. 
However, the interior of the ring ($r\lesssim 20$~au) is filled by dust.
Thus, this structure is observed as a gap structure in the dust emission.
In the case of the filled dust disk ($\alpha_{\rm MRI}=10^{-5}$, $\Cw=10^{-2}$, St=$10^{-3}$), the surface density of the dust has no remarkable structure.
On the other hand, the gas is quickly removed by the wind mass loss resulting in no or quite small amount of gas in the disk.
In contrast to this case, there is no structure in the gas surface density and 
only few amount of dust remains in the case of the dust poor disk ($\alpha_{\rm MRI}=10^{-2}$, $\Cw=10^{-5}$, St=1).
This small dust surface density is due to the efficient radial drift of the dust.
Finally, in the case of normal filled disk ($\alpha_{\rm MRI}=10^{-2}$, $\Cw=10^{-5}$, St=$10^{-3}$), there is no small scale structure in both gas and dust surface density.
In this case, the dust-to-gas mass ratio is about 0.01, which is the same as that of the infalling envelope.
The diversity of the disk structure can be attributed to the difference of timescales of viscosity, wind mass loss, and dust radial drift, as described in the next section.

\subsection{Comparison of Timescales}
The formation of various disk structures can be understood by comparing three important timescales: the timescales of the viscous diffusion, the wind mass loss and the radial drift of the dust.
We estimate the viscous timescale by using the standard viscous disk with Kepler rotation.
The evolution of the surface density only due to the viscosity is given by
\begin{eqnarray}
  \frac{\partial \Sigma}{\partial t}
&=&\frac{3}{r} \frac{\partial}{\partial r}
\left(
\sqrt{r}\frac{\partial}{\partial r}
 \sqrt{r}\Sigma \nu
\right)\nonumber\\
&=&3(n_{\Sigma}+n_{\nu}+0.5)(n_{\Sigma}+n_{\nu})
\Sigma\nu r^{-2},
\label{eq:time_vis}
\end{eqnarray}
where $n_{\Sigma}=d\ln\Sigma/d\ln r$ and $n_{\nu}=d\ln\nu/d\ln r$.
We now consider the location close to the local maximum of the pressure, where dust particles are expected to accumulate.  
At such radius, $p\propto \Sigma T_{\rm eq}/H =$const, where $H=\cs/\Omega$ is the disk scale height.
Thus, we obtain $\Sigma\propto H/T_{\rm eq}$ and $\nu\Sigma\propto (\cs^2/\Omega)(H/T_{\rm eq})\propto T_{\rm eq}^{0.5}\Omega^{-2} \propto r^{2.8}$.
This gives $n_{\Sigma}+n_{\nu}=2.8$ and the viscous timescale 
\begin{eqnarray}
 t_{\rm vis} &\sim& \frac{r^2}{3(n_{\Sigma}+n_{\nu}+0.5)(n_{\Sigma}+n_{\nu})\nu} \nonumber \\
& =& 8\times 10^4\ [{\rm yr}] 
\left(\frac{r}{10{\rm au}}\right)^{13/14}
\left(\frac{\alpha}{10^{-3}}\right)^{-1}.
\label{eq:tvis}
\end{eqnarray}
The wind timescale is 
\begin{equation}
 t_{\rm wind} = \frac{\Sigma}{{\dot \Sigma}_{\rm wind }} 
=5\times10^4[{\rm yr}] 
\left(\frac{r}{10{\rm au}}\right)^{3/2}
\left(\frac{\Cw}{10^{-4}}\right)^{-1}.
\label{eq:twind}
\end{equation}
The radial drift timescale is 
\begin{eqnarray}
 t_{\rm drift}&=&\frac{r}{v_{\rm drift}}
 =\frac{1}{\Omega \St}\left(\frac{\cs}{r\Omega}\right)^{-2}\left|\frac{d\ln p}{d\ln r}\right|^{-1}\nonumber\\
&=&  6.6\times10^3[{\rm yr}]
\label{eq:tdrift}
  \left(\frac{r}{10 [\rm au]}\right)^{13/14}
  \left(\frac{\St }{10^{-1}}\right)^{-1},
\end{eqnarray}
where we assume St and dust-to-gas mass ratio are much smaller than unity and the gas surface density is given in such a way that Toomre's $Q$-parameter is unity (Equation (\ref{eq:sigma_Q1})).

\begin{figure*}[tb]
  \begin{minipage}{0.3\hsize}
   \begin{center}
    \includegraphics[width=6cm]{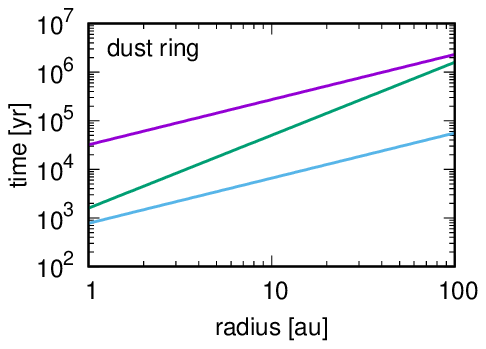}
   \end{center}   
  \end{minipage}
  \begin{minipage}{0.3\hsize}
   \begin{center}
    \includegraphics[width=6cm]{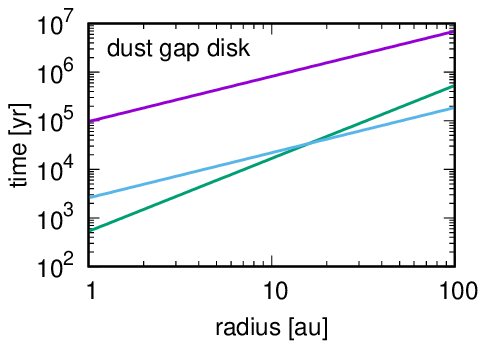}
   \end{center}   
  \end{minipage}
  \begin{minipage}{0.3\hsize}
   \begin{center}
    \includegraphics[width=6cm]{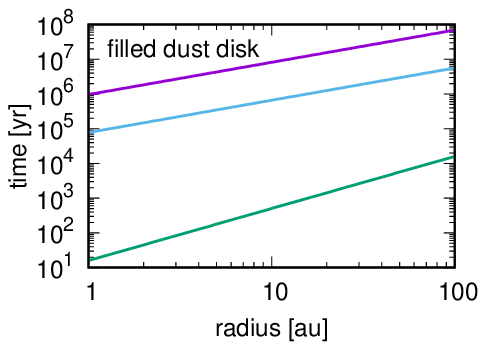}
   \end{center}   
  \end{minipage}
  \begin{minipage}{0.3\hsize}
   \begin{center}
    \includegraphics[width=2cm]{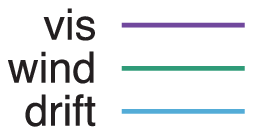}
   \end{center}   
  \end{minipage}
  \hspace{19pt}
  \begin{minipage}{0.3\hsize}
   \begin{center}
    \includegraphics[width=6cm]{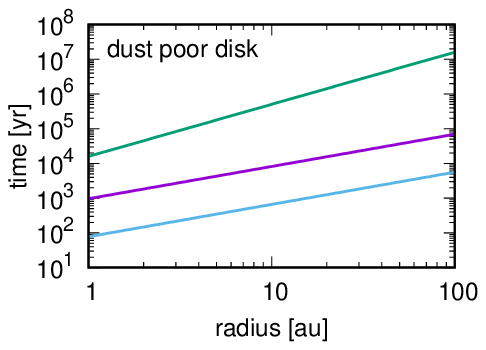}
   \end{center}   
  \end{minipage}
  \hspace{10pt}
  \begin{minipage}{0.3\hsize}
   \begin{center}
    \includegraphics[width=6cm]{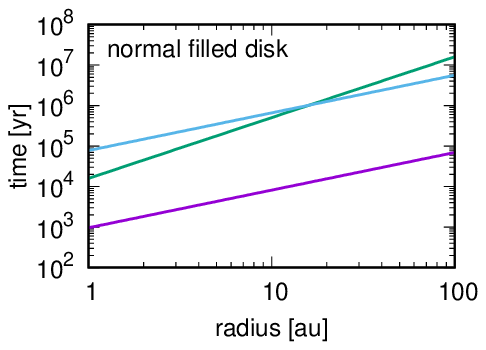}
   \end{center}   
  \end{minipage}
\caption{Comparison of the time scales for the dust ring ($\alpha_{\rm MRI} = 3\times 10^{-4},\ \Cw=10^{-4}$ and St=0.1), the dust gap disk ($\alpha_{\rm MRI} = 10^{-4},\ \Cw=3\times 10^{-4}$ and St=0.03), the filled dust disk ($\alpha_{\rm MRI} = 10^{-5},\ \Cw=10^{-2}$ and St=$10^{-3}$), the dust poor disk ($\alpha_{\rm MRI} = 10^{-2},\ \Cw=10^{-5}$ and St=1), and normal filled disk ($\alpha_{\rm MRI} = 10^{-2},\ \Cw=10^{-5}$ and St=$10^{-3}$).  The purple, green and blue lines show the timescales of the viscous diffusion, wind mass loss and the radial drift of the dust, respectively.}
\label{fig:time_st1}
 \end{figure*}

Figure \ref{fig:time_st1} shows the timescales of the viscous diffusion, the wind mass loss and the radial drift of the dust for five parameter sets explained above.
\begin{itemize}
 \item Dust ring disk: The viscous timescale $t_{\rm vis}$ is larger than the wind timescales $t_{\rm wind}$. 
Thus, the time evolution of the gas surface density is mainly determined by the disk wind after the wind mass loss starts. 
The wind timescale around 10-20 au is about $10^5$ yr. 
Thus, the wind mass loss makes the pressure maximum at $\sim$ 10 au at $t\sim 0.6$~Myr, which is $\sim 0.3$~Myr after the time when the wind mass loss starts.
Since the wind timescale is larger for larger radius, the radius of the pressure maximum moves outward with time.
This makes the ring move outward as shown in Figure \ref{fig:sigma_st1} and \ref{fig:ring_timeevo1}.
The drift timescale $t_{\rm drift}$ is comparable to or smaller than the wind timescale.
Thus, the dust can concentrate on the pressure maximum formed due to the disk wind and the ring structure is formed.

\item Dust gap disk: In this case, the time evolution of the gas disk is dominated by the wind mass loss because $t_{\rm vis} > t_{\rm wind}$. 
In the inner part of the disk ($r\lesssim 20$~au), the gas in the disk is removed before dust particles concentrate at the location of pressure maximum since $\tdrift>\twind$.
As a result, the dust inner disk is formed in $r\lesssim$ 20 au.
On the other hand, $\tdrift<\twind$ is satisfied at outer radii ($r\gtrsim$ 20~au). 
This is similar to the case of the ring structure formation described above.
Therefore, the ring like structure is also formed in $r\gtrsim 20$ au in this case. 
The gap structure consists of the inner dust disk and the outer ring structure.
\item Filled dust disk: In the case that $\twind<(\tvis,\ \tdrift)$ is satisfied in the whole disk, the dust disk without gas is formed since the gas is removed by the wind very rapidly.
This structure corresponds to the inner region of the gap structure.
\item Dust poor disk: When $\tvis<\twind$, the hole structure of the gas disk is not formed by the wind mass loss.
Since the radial drift is faster than the viscous evolution ($\tdrift<\tvis$), the dust surface density is much smaller than that of the gas, resulting in the dust poor disk.
\item Normal filled disk: In this case, neither the gas nor the dust hole structures are formed since $\tvis<\twind$.
However, in contrast to the case of dust poor disk, $\tdrift<\tvis$ is satisfied in the disk.
Thus, the radial drift of the dust is not significant and the dust-to-gas mass ratio is almost same as that of the infall envelope.
\end{itemize}

\begin{figure*}[tb]
\begin{center}
\includegraphics[width=14cm]{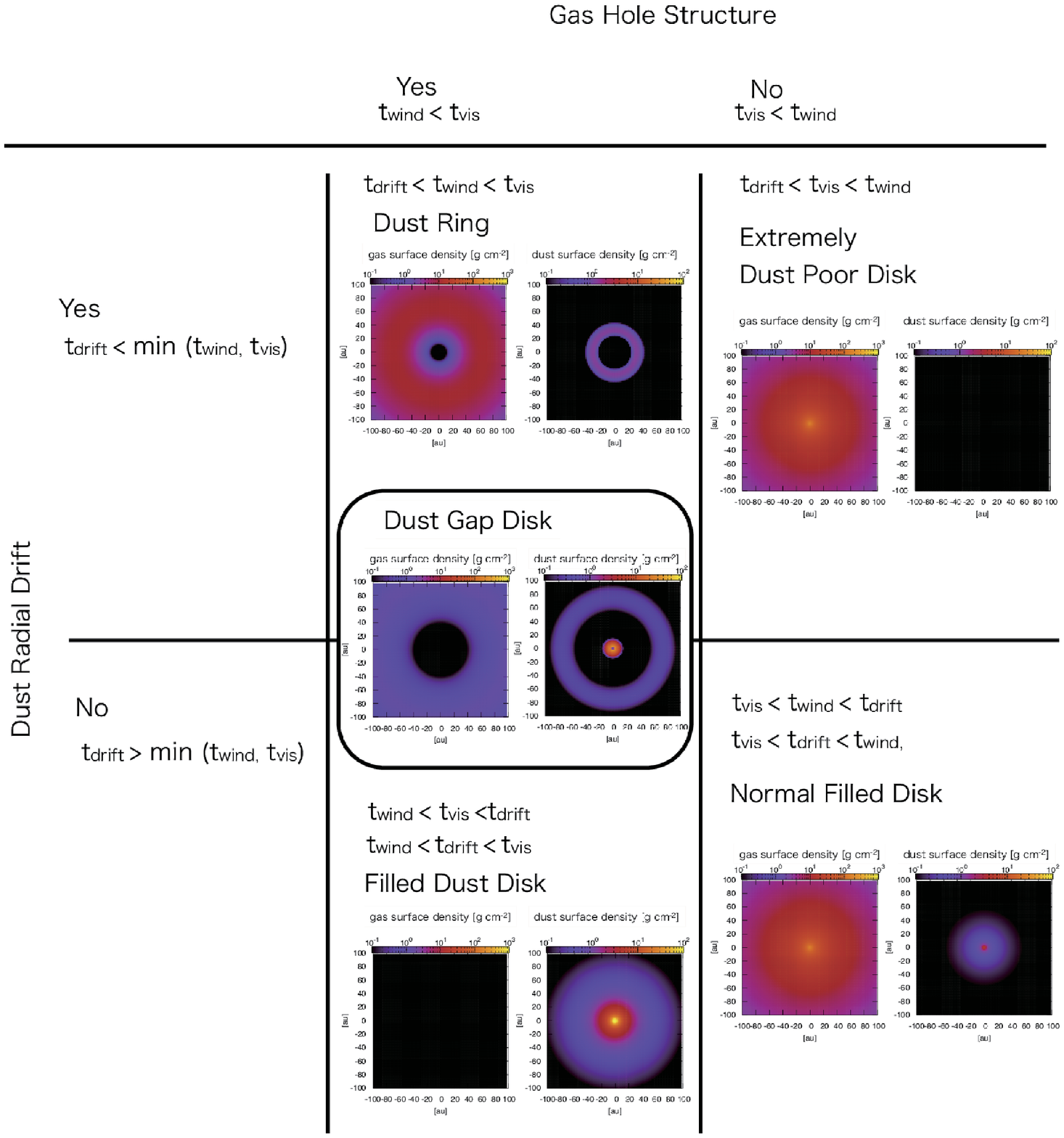}
\vspace{-5cm}
\caption{Relation between the timescales and the disk structure. Color maps show the surface density distributions of the gas and the dust for each disk structure at $t=1.27$ Myr.
}
\label{fig:type_table}		    
\end{center}
\end{figure*}

The relation between the timescales and the disk structure is summarized in Figure \ref{fig:type_table}.

\subsection{Parameter dependence of ring structure}
In this section, we present the results of parameter study, where we have varied $\alpha_{\rm MRI}$, $\Cw$, and St.
We investigate how these parameters are related to the disk morphology.  We focus on the ring radius, the ring width, the dust surface density at the ring, existence of the inner dust disk interior to the ring (cf. the dust disk in $r\lesssim 20$ au in left top panel of Figure \ref{fig:sigma_type} (dust gap disk)), and dust-to-gas mass ratio in the ring structure.
The ring radius is defined by the radius of the local maximum of the dust surface density.
The ring width is given by the full width of the half maximum of the dust surface density distribution.
We regard the local maximum of the dust surface density as a ring when the inner radius of the ring is larger than the radius of the inner boundary.
We only investigate the ring whose radius is larger than 3 au, because the surface density distribution at the innermost region of the computational domain may be affected by the inner boundary condition.

\begin{figure}[tb]
 \includegraphics[width=9cm]{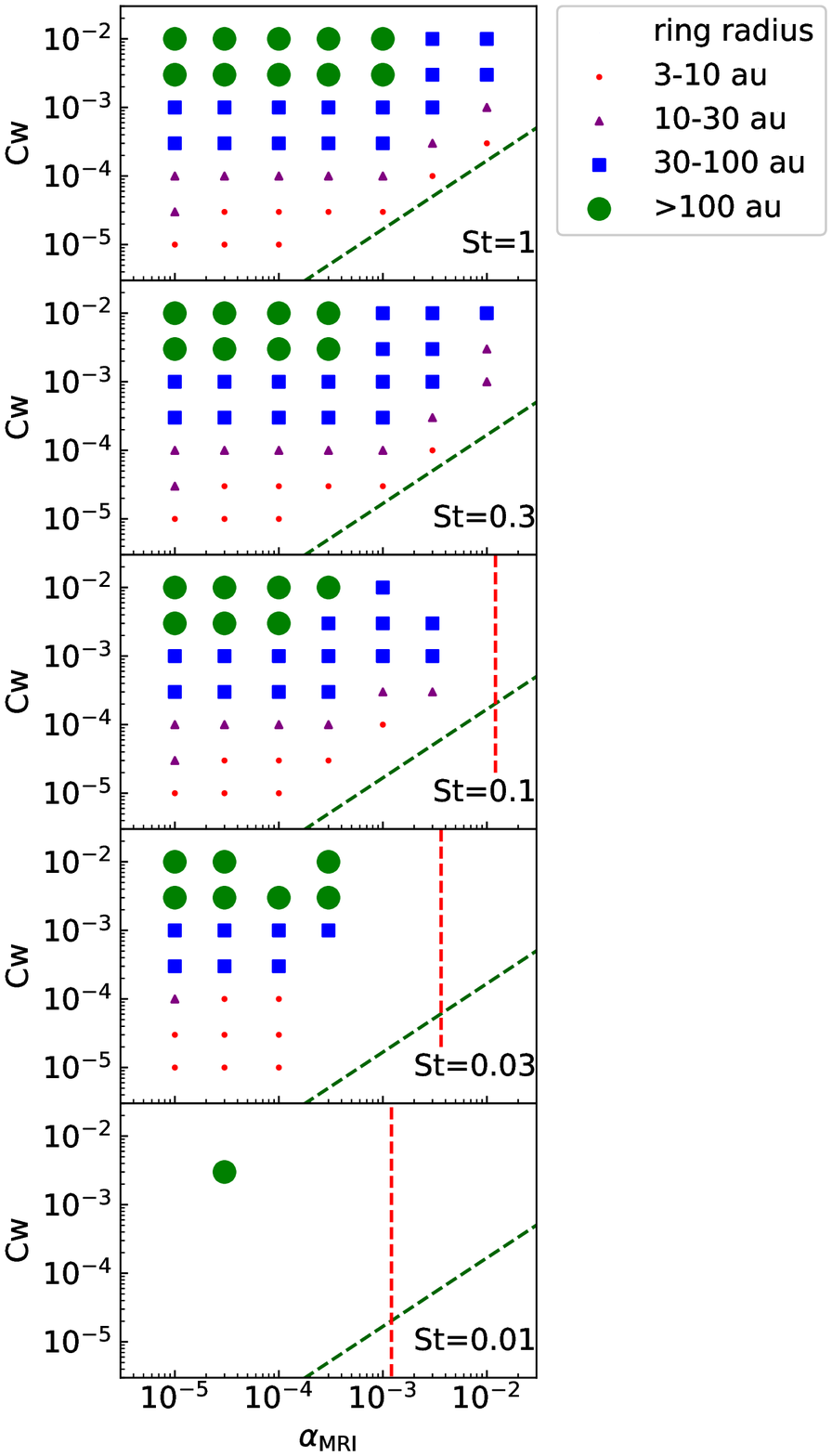}
 \caption{Ring radius at $t=0.6$~Myr. Horizontal axis is $\alpha_{\rm MRI}$ and vertical axis is $\Cw$. Each panels shows the result with $\St=1,\ 0.3,\ 0.1,\ 0.03,\ 0.01$ from top to bottom.
Small red circles, purple triangles, blue squares, and large green circles indicate the ring radius of 3-10 au, 10-30 au, 30-100 au, and larger than 100 au, respectively.
Green dashed lines show the relation of $\alpha_{\rm MRI}$ and $\Cw$ satisfying $t_{\rm vis}=t_{\rm wind}$ at the inner boundary.
The region below this line corresponds to $t_{\rm vis}<t_{\rm wind}$.
Red dashed lines show the relation of $\alpha_{\rm MRI}$ and $\St$ satisfying $t_{\rm vis} = t_{\rm drift}$
.
The left of this line corresponds to $t_{\rm vis} > t_{\rm drift}$.}
\label{fig:bbl_rad}
\end{figure}

Figure \ref{fig:bbl_rad} shows the ring radius at $t=0.6$~Myr for $10^{-5}\leq\alpha_{\rm MRI}\leq10^{-2}$, $10^{-5}\leq\Cw\leq10^{-2}$ and $10^{-2}\leq \St\leq1$.
The green dashed lines indicate the parameters where $\alpha_{\rm MRI}$ and $\Cw$ satisfy $\tvis=\twind$ at the inner boundary.
The region under this line corresponds to $t_{\rm vis}<t_{\rm wind}$.
In this region, the viscous diffusion prevents the inner hole formation in the gas disk by the wind mass loss.
As a result, the ring structure is not formed.
This disk structure corresponds to the  dust poor disk or the normal filled disk.
The red dashed lines indicate the parameters where $\alpha_{\rm MRI}$ and St satisfy $\tvis=\tdrift$.
The parameters right of the red lines and above the green lines correspond to the case where the timescale of the wind mass loss is smaller than the radial drift of the dust ($t_{\rm wind} < t_{\rm vis} < t_{\rm drift}$).
If $\twind<\tdrift$, ring structures do not form since dust particles cannot accumulate to the gas pressure maximum before gas is lost by the wind.  Therefore, this parameter space corresponds to the filled dust disk.
For $\St=0.01$, the radial drift timescale is of order of $10^5$ yr in the disk.
Thus, the ring structure is not formed in most of the parameters of $\alpha_{\rm MRI}$ and $\Cw$.
It is noted that the case when $\alpha_{\rm MRI}=10^{-4}$, $\Cw=10^{-2}$ and St=0.03 is categorized as no ring structure, despite we see ring structures in other parameters around them.  This is partly due to our setting of threshold values to determine ``ring structure''.  We require that
the dust surface density contrast in the ring is larger than factor of two.
Around the parameter region, dust surface density contrast is about factor of two, which is close to our threshold value.  We do see weak ring structures are formed when $\alpha_{\rm MRI}=10^{-4}$, $\Cw=10^{-2}$ and St=0.03. 
The same is true with parameters around $\alpha_{\rm MRI}=3\times10^{-5}$, $\Cw=10^{-3}$ and St=0.01.

\begin{figure}[tb]
 \includegraphics[width=9cm]{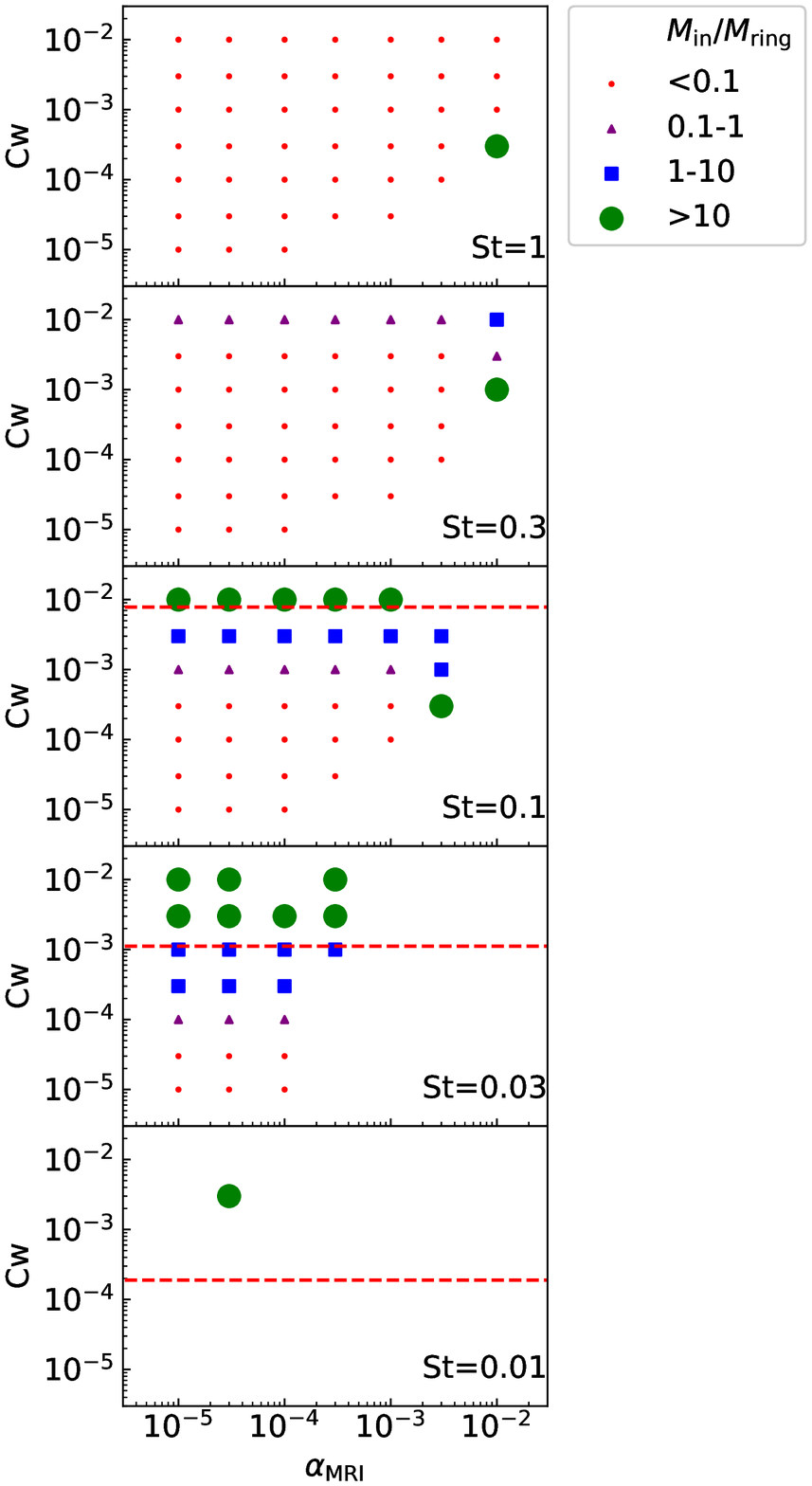}
 \caption{Dust mass ratio of the inner ring and in the ring at $t=0.6$~Myr. Horizontal axis is $\alpha_{\rm MRI}$ and vertical axis is $\Cw$. Each panels shows the result with $\St=1,\ 0.3,\ 0.1,\ 0.03,\ 0.01$ from top to bottom.
Small red circles, purple triangles, blue squares, and large green circles indicate the ring radius of 0.1, 0.1-1, 1-10, and larger than 10, respectively.
The red dashed lines shows the critical $\Cw$ to make the inner hole obtained from the comparison of $t_{\rm wind}$ and $t_{\rm drift}$.
}
\label{fig:bbl_Min}
\end{figure}

In Figure \ref{fig:bbl_rad}, the ``ring radius'' is measured in two kinds of dust distribution.  One is the ``dust ring disk'' and the other is the ``dust gap disk''.  For the dust gap disk, the bump of the surface density at the outer edge of the gap (e.g, the dust structure in $r\gtrsim 20$~au in the top left panel of Figure \ref{fig:sigma_type}) is  classified as the ring structure. 
To distinguish the ring structure and the dust gap disk, we show the ratio of the dust mass inside the ring and the total ring dust mass in Figure \ref{fig:bbl_Min}.
For the dust with small Stokes number, the radial drift timescale is large.
When the drift timescales is larger than the wind timescale, the expansion of the inner hole of the gas disk due to wind mass loss is faster than the redial drift of the dust.
As a result, the inner hole is not formed in the dust distribution for small St.
The critical value of $\Cw$ as to whether the dust inner disk remains or not can be estimated from the comparison of $t_{\rm wind }$ and $t_{\rm drift}$ as follows.
The ring radius is estimated by the wind timescale given by Equation (\ref{eq:twind}).
Since the wind mass loss start at $t\sim 0.27$~Myr and we focus on the ring structure at $t=0.6$~Myr here, the relation of the ring radius and $\Cw$ is obtained from Equation (\ref{eq:twind}) by substituting 0.3~Myr for $t_{\rm wind}$,
\begin{equation}
\left(\frac{r}{10\ {\rm au}}\right) = 3.3\left(\frac{\Cw}{10^{-4}}\right)^{2/3}.
\label{eq:rad-Cw_relation}
\end{equation}
Then, substituting this radius in Equation (\ref{eq:tdrift}), we obtain the drift timescale at the ring radius.
To make the inner hole in the dust disk, this timescale is required to be smaller than $t_{\rm wind }=0.3$~Myr.
As a result, we obtain the critical $\Cw$
\begin{equation}
 \Cw = 7.9\times10^{-3} \left(\frac{\St}{10^{-1}}\right)^{21/13}.
\label{eq:critical_cw}
\end{equation}
The red dashed lines in Figure \ref{fig:bbl_Min} show this critical $\Cw$.
This figure shows that the critical $\Cw$ estimated here roughly traces the mass ratio $M_{\rm in}/M_{\rm ring}\sim 10$.
The mass ratio $\sim 1$ corresponds to $\Cw$ that is an order of magnitude smaller than the critical $\Cw$.
The disk structure with large amount of dust inside the ring ($M_{\rm in}/M_{\rm ring}\gtrsim 1$) corresponds to the dust gap disk.

\begin{figure}[tb]
 \includegraphics[width=9cm]{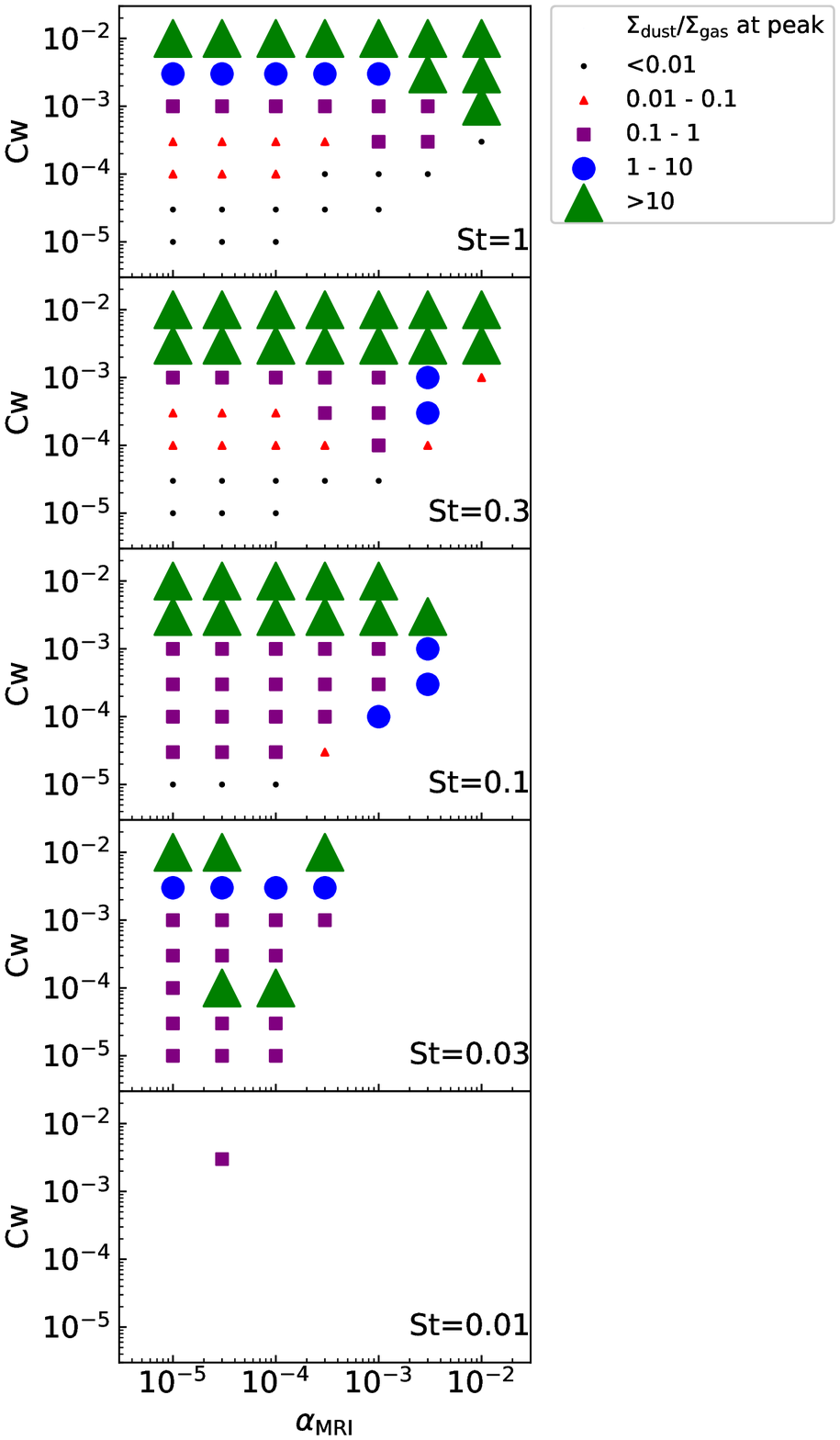}
 \caption{Dust-to-gas mass ratio at the ring radius at $t=0.6$~Myr. Horizontal axis is $\alpha_{\rm MRI}$ and vertical axis is $\Cw$. Each panels shows the result with $\St=1,\ 0.3,\ 0.1,\ 0.03,\ 0.01$ from top to bottom.
Small black circles, red triangles, purple squares, large blue circles and large green circles indicate the dust-to-gas mass ratio of smaller than 0.01, 0.1-1, 1-10, and larger than 10, respectively.
}
\label{fig:bbl_ratio}
\end{figure}

The dust-to-gas mass ratio is the other important value because it is strongly related to the planetesimal formation in the ring.
Figure \ref{fig:bbl_ratio} shows the dust-to-gas mass ratio at the ring radius.
This figure shows that the ratio is large for large $\Cw$ because the gas surface density decreases rapidly when the wind mass loss is efficient. 
The ring radius is large when $\Cw$ is large, and therefore, dust-to-gas mass ratio is large when the ring radius is large.
The dust-to-gas mass ratio is large for small Stokes number.
This is because the radial drift of the dust with small Stokes number is not efficient so that the dust does not drift towards the central star before the wind mass loss starts. 
As a result, the total mass of the dust in the disk is large for small Stokes number, resulting in the large dust-to-gas ratio at the ring.

\begin{figure}[tb]
 \includegraphics[width=9cm]{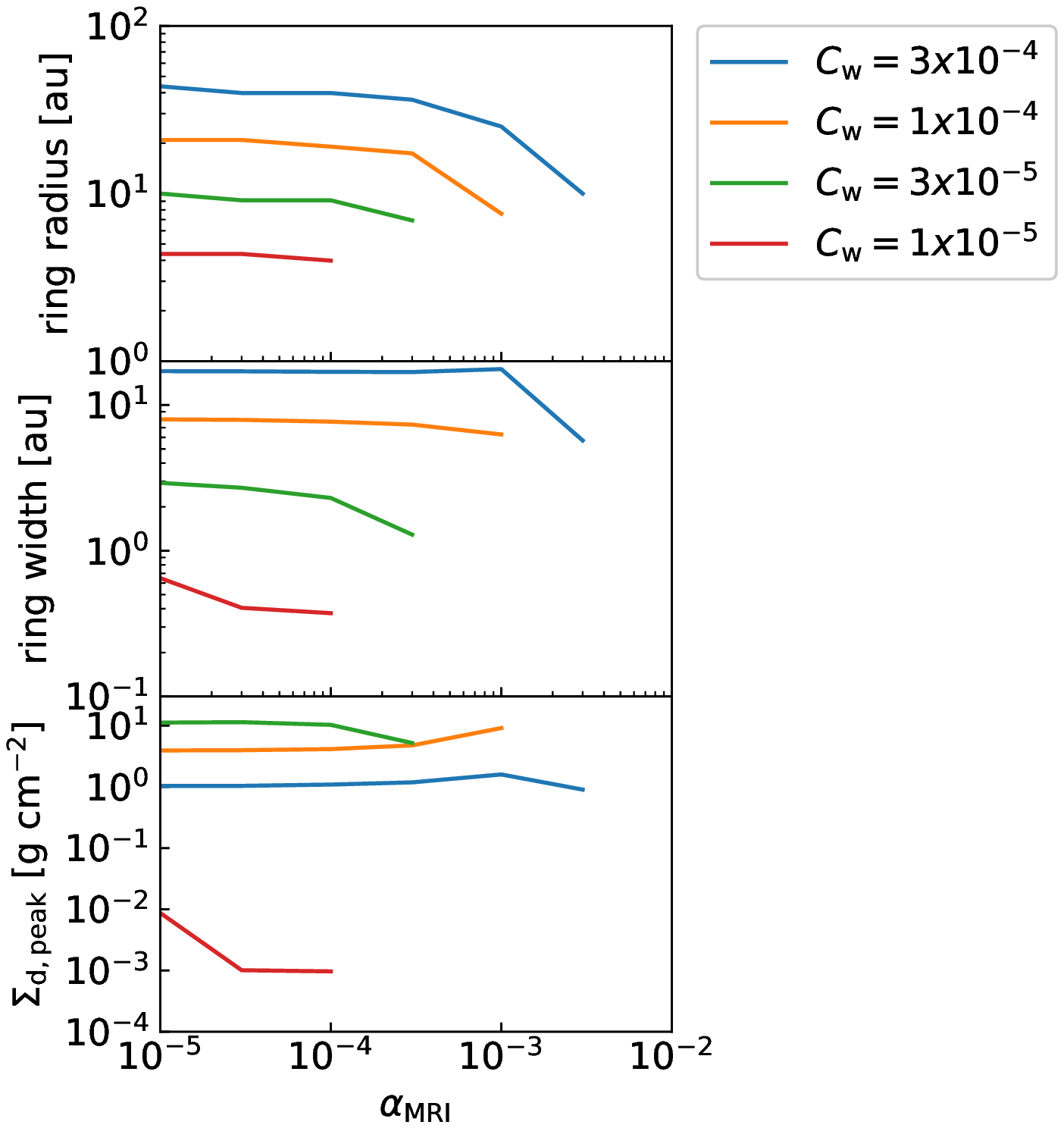}
\caption{Ring radius, width and maximum surface density of dust with $\St=0.1$ at $t=0.6$~Myr from top to bottom. 
}
\label{fig:ring-alpha_St01}
\end{figure}

Finally, we investigate how the ring radius, width, and the maximum surface density depends on input parameters.
Figure \ref{fig:ring-alpha_St01} shows these values for various $\alpha_{\rm MRI}$ and $\Cw$ with St=0.1 at $t=0.6$~Myr.
The top panel of Figure \ref{fig:ring-alpha_St01} shows the parameter dependence of the ring radius.
For $\alpha_{\rm MRI}\lesssim 3\times 10^{-4}$, the ring radius depends only on $\Cw$ because the disk evolution is dominated by the disk wind.
For $\alpha_{\rm MRI} \gtrsim 3\times 10^{-4}$, the viscous diffusion is efficient and $t_{\rm wind}<t_{\rm vis}$ is satisfied only in the inner region.
As a result, the ring radius is small for large $\alpha_{\rm MRI}$.
The middle panel of Figure \ref{fig:ring-alpha_St01} shows the width of the ring.
The ring width is of the order of 10 au when $\Cw=10^{-4}$ and $3\times10^{-4}$. 
For $\Cw\lesssim 3\times10^{-5}$, the ring width is of order of 1 au because the large wind timescale allows gas to remain in the disk for long time and the dust particles concentrate strongly at the location of pressure maximum.
The bottom panel of Figure \ref{fig:ring-alpha_St01} shows the maximum surface density of the dust in the ring.
For $\Cw=10^{-5}$, the wind timescale is larger than the drift timescale so that only weak gas pressure maximum is formed and the dust surface density decreases due to the radial drift to the central star.
As a result, the dust surface density with $\Cw=10^{-5}$ is much smaller than those of the other cases.

\begin{figure}[tb]
 \includegraphics[width=9cm]{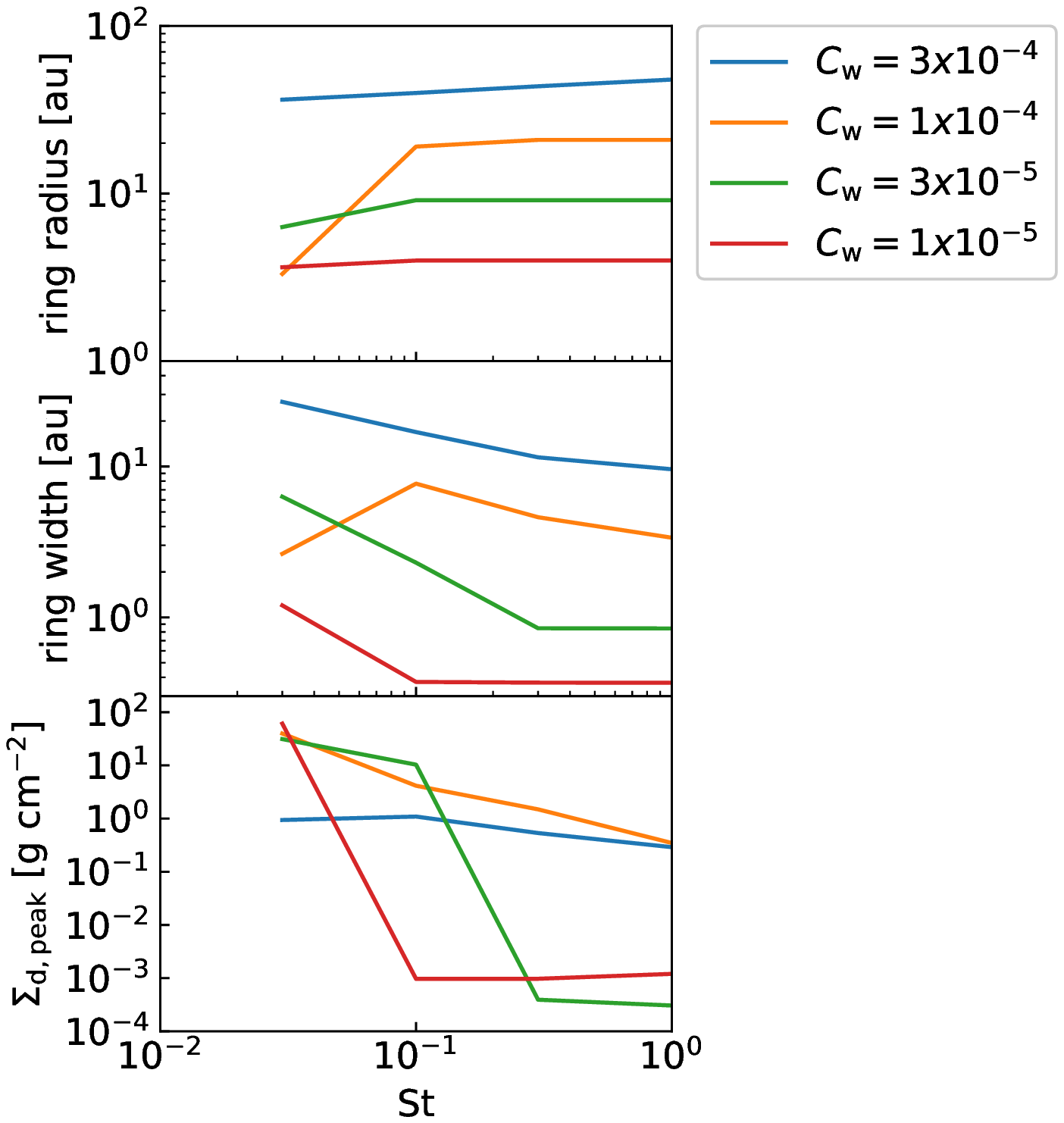}
\caption{Ring radius, width and maximum surface density of with $\alpha_{\rm MRI}=10^{-4}$ at $t=0.6$~Myr from top to bottom. 
}
\label{fig:ring-St_alpha1em4}
\end{figure}

Figure \ref{fig:ring-St_alpha1em4} shows the ring radius, the width and the maximum surface density of dust for various $\Cw$ and St with $\alpha_{\rm MRI}=10^{-4}$ at $t=0.6$~Myr.
The top panel of Figure \ref{fig:ring-St_alpha1em4} is the ring radius.
For $\St\lesssim 3\times10^{-2}$, the dust strongly couples with the gas and cannot concentrate into the pressure maximum.
Thus, no ring structure is formed in the disk.
For $\St\gtrsim 10^{-1}$, the radius of the ring does not strongly depend on St because the radius of the pressure maximum depends mainly on the wind efficiency and viscous diffusion timescale.
The middle panel of Figure \ref{fig:ring-St_alpha1em4} shows the ring width. For $\St>10^{-1}$, the ring width decreases with St increases.
This tendency is understood as a result of the rapid concentration of the dust in the pressure maximum.
The bottom panel of Figure \ref{fig:ring-St_alpha1em4} shows the maximum surface density in the ring.
For large Stokes number, the radial drift timescale is small and the dust surface density decreases before the wind mass loss stats.
As a result, the total dust-to-gas mass ratio and the ring surface density is small for the dust with large Stokes number.

\section{Discussion}
\label{discussion}

\subsection{Dependence on the angular velocity of the cloud cores}
\label{jcore}
\begin{figure}[tb]
\begin{tabular}{c}
\includegraphics[width=9cm]{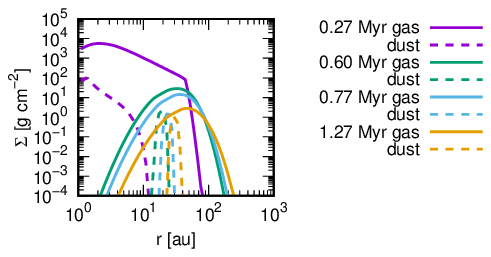}
 \\
\includegraphics[width=9cm]{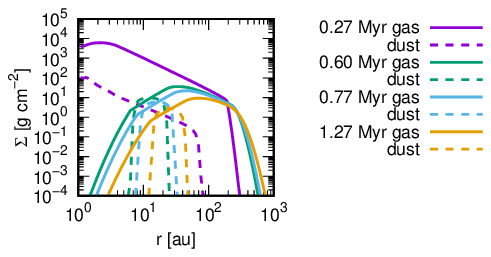}
\end{tabular}
\caption{Time evolution of the surface density of the gas and dust with $\alpha_{\rm MRI} = 3\times 10^{-4},\ \Cw=10^{-4}$ and $\St=0.1$. The top and bottom panels show the results with $\Omega_{\rm core}=0.2\ {\rm and}\ 0.5[{\rm km s^{-2} pc^{-1}}]$, respectively.The solid lines shows the gas surface density and the dotted lines shows the dust surface density.}
\label{fig:sigma_j}
\end{figure}

In this section, we discuss the effect of the angular velocity of the cloud core, which we have fixed at $\Omega_{\rm core}=0.3 \ {\rm [km\ s^{-1}\ pc^{-1}]}$.
The angular momentum affects mainly the dust mass in the disk at the time when the wind mass loss starts.
Figure \ref{fig:sigma_j} shows the evolution of the surface density with $\alpha_{\rm MRI} = 3\times 10^{-4},\ \Cw=10^{-4}$ and $\St=0.1$, which are same as those of Figure \ref{fig:sigma_st1}.
The top and bottom panels are the result with $\Omega_{\rm core} = 0.2\ {\rm [km\ s^{-1}\ pc^{-1}]}$ and $0.5\ {\rm [km\ s^{-1}\ pc^{-1}]}$, respectively.
For $\Omega_{\rm core} = 0.2\ {\rm [km\ s^{-1}\ pc^{-1}]}$, the maximum centrifugal radius is about 10 au and the infalling envelope accrete onto the inner radius of the disk.
Thus, the radius of the gas and dust is smaller than that with $\Omega_{\rm core} = 0.3\ {\rm [km\ s^{-1}\ pc^{-1}]}$.
As a result, the total dust mass in the disk is also small.
Thus, the resultant ring surface density is smaller than that with $\Omega_{\rm core} = 0.3\ {\rm [km\ s^{-1}\ pc^{-1}]}$.
On the other hand, the ring radius is almost the same as that with $\Omega_{\rm core} = 0.3\ {\rm [km\ s^{-1}\ pc^{-1}]}$, because it is mainly determined by the wind timescale.
For the case with $\Omega_{\rm core} = 0.5\ {\rm [km\ s^{-1}\ pc^{-1}]}$, the large gas disk is formed.
The total dust mass is also larger than that with $\Omega_{\rm core} = 0.3\ {\rm [km\ s^{-1}\ pc^{-1}]}$.
Because of the large dust mass, dust-to-gas mass ratio becomes about unity at $t=0.6$~Myr.
In this case, the back reaction from dust to gas prevents the radial drift of the dust.
As a result, the width of the ring is larger than that of $\Omega_{\rm core} = 0.3\ {\rm [km\ s^{-1}\ pc^{-1}]}$.
In this case, the gas surface density distribution is changed by the back reaction from dust to gas.
Thus, the location of pressure maximum moves inward and the ring radius is slightly smaller than that of $\Omega_{\rm core} = 0.3\ {\rm [km\ s^{-1}\ pc^{-1}]}$.

\subsection{Constant dust radius}
\label{a_dust}
\begin{figure}[tb]
\includegraphics[width=9cm]{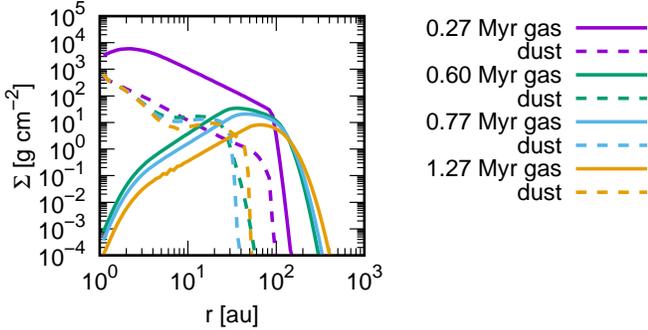}
\caption{Time evolution of the surface density of the gas and dust with $\alpha_{\rm MRI} = 3\times 10^{-4},\ \Cw=10^{-4}$ and $a=1$ mm. The solid lines shows the gas surface density and the dotted lines shows the dust surface density.}
\label{fig:sigma_a1}
\end{figure}

\begin{figure}[tb]
\includegraphics[width=9cm]{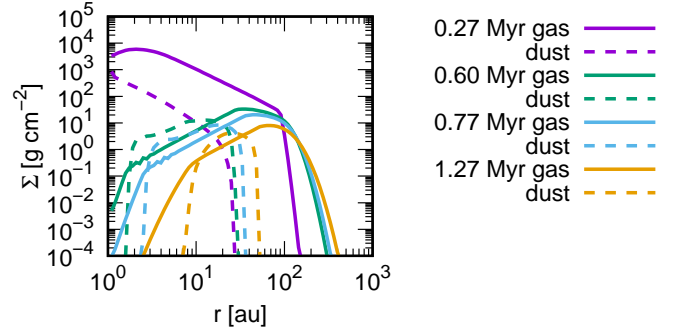}
\caption{Time evolution of the surface density of the gas and dust with $\alpha_{\rm MRI} = 3\times 10^{-4},\ \Cw=10^{-4}$ and $a=3$ cm. The solid lines shows the gas surface density and the dotted lines shows the dust surface density.}
\label{fig:sigma_a2}
\end{figure}

In this paper, we have fixed the Stokes number of dust particles within the disk for simplicity, as motivated by previous study of grain growth.  The actual dust properties within the disk is largely unknown, and therefore we have also tried another extreme, where the size of dust particles does not change within the disk.
Figure \ref{fig:sigma_a1} shows the time evolution of the surface density of the gas and the dust with $\alpha_{\rm MRI} = 3\times 10^{-4},\ \Cw=10^{-4}$ and $a=1$ mm.
In this case, the dust particles are tightly coupled with the gas. 
In the inner region, the Stokes number is $\St\sim 10^{-5}$. 
Thus, the dust-to-gas mass ratio is about $\sim0.01$ in the disk at $t\sim 0.27$~Myr.
The wind mass loss makes the inner hole in the gas disk in $t>0.27$~Myr.
However, the dust inner hole is not formed because the dust is coupled too tightly with the gas to drift in the disk and to concentrate at the pressure maximum of the gas disk.
This is even in the case with the large dust radius.
Figure \ref{fig:sigma_a2} shows the evolution of the surface densities of the gas and the dust with $\alpha_{\rm MRI} = 3\times 10^{-4},\ \Cw=10^{-4}$ and $a=3$ cm, which corresponds to St$\sim 3\times10^{-4}$ at 1 au at $t\sim 0.27$~Myr.
In this case, the coupling between the gas and the dust is weaker than the previous case since the dust size is larger.
However, the inner hole of the dust is not formed even in this case although the dust surface density in $r\lesssim 10$ au is smaller than the previous case.
We have checked for other dust sizes and found that 
the dust inner hole is not formed when the protostar age is $\sim 0.5$~Myr if we assume the constant dust radius $a<3$ cm.

\subsection{Treatment of the disk wind in Class 0, I YSOs}
\label{windmodel}
In our model, the wind mass loss starts after the mass accretion from the cloud core onto the disk terminates for simplicity.
In reality, however, the existence of the outflow in Class 0, I phase is reported by both observations \cite[e.g.][]{2006ApJ...646.1070A,2013A&A...556A..76V} and numerical simulations \cite[e.g.][and references therein]{1998ApJ...502L.163T,2012PTEP.2012aA307I}.
Since the outflow transports the mass and angular momentum of the disk, the outflow in the early star formation phase will decrease the total mass and angular momentum of the Class II star-disk system.

The effect of the outflow in Class 0, I phase on the disk surface density distribution is not yet fully understood.
The numerical simulations for the star and disk formation have shown that the disk with outflow is gravitationally unstable \cite[cf.][]{2010ApJ...718L..58I}.
The surface density profile of such a disk is expected to satisfy $Q\sim 1$.
Thus, the disk structure before the infall stops obtained from our model (e.g. upper panel of Fig \ref{fig:sigma_st1}) might mimic that with the outflow.
Even when the outflow changes the surface density profile of the gas and the dust at the end of the infall, 
we expect that the classification of the disk morphology at the early phase of disk formation may be applicable at least qualitatively
because it depend only on the timescales of the viscous diffusion, the wind mass loss and the radial drift of the dust.

To confirm the validity of our model and obtain the realistic processes of the disk structure formation, we need to develop  the model treating the outflow in Class 0, I phase consistently.
This involves detailed comparison between the 1D model and the state-of-the-art 3D numerical simulations, which we shall explore in future work.

\subsection{Comparison with the observation of WL~17}
\label{WL17}
In this subsection, we discuss the parameter set of $\alpha_{\rm MRI}$, $\Cw$, and $\St$ suitable to explain the observed structure in the disk around WL~17.
\cite{2017ApJ...840L..12S} presented observations as well as a simple model for this object.  In the following discussion, we quote their values as demonstration.
The observations show that the ring radius is 10-20 au.
According to Figure \ref{fig:bbl_rad} (see also Figure \ref{fig:ring-alpha_St01}), it corresponds to the result with $\Cw\sim 10^{-4}$.
To make the ring structure with  $\Cw\sim 10^{-4}$, it is also required that $\alpha_{\rm MRI} \lesssim 10^{-3}$.
Since dust emission from WL~17 shows a ring-like structure, the dust and gas should be reasonably decoupled and the Stokes number larger than about 0.03 is preferred.
However, the Stokes number should not be too large considering the disk mass (see Figure \ref{fig:ring-St_alpha1em4}).
Since the disk mass of WL~17 is about 0.05 $M_{\odot}$ and the disk area is $\sim 300\pi\ [{\rm au^2}]$, the mean surface density of dust is 
$10^{-2} \times 0.05M_{\rm \odot}/(300\pi [{\rm au^2}]) \sim 5\ [{\rm g\ cm^{-2}}]$.
Comparing this surface density with results of model calculations, Stokes number maybe lower than 0.3.
The outflow of the WL~17 is also observed by the $^{12}$CO line emission in \cite{2013A&A...556A..76V}.
The estimated mass and timescale of the observed outflow are $\sim 10^{-4} M_{\rm \odot}$ and $\sim 10^3$ yr. 
Thus, the mass loss rate is $\sim 10^{-7}\ [M_\odot\ {\rm yr}^{-1}]$.
As shown in Figure \ref{fig:Mdotwind}, the mass loss rate is of the order of $\sim 10^{-7}\ [M_\odot\ {\rm yr}^{-1}]$ at $t\sim 0.5$~Myr when $\alpha_{\rm MRI}=3\times10^{-4}$, $\Cw=10^{-4}$, St=0.1.
This mass loss rate is also consistent with the value obtained form observations.
Based on the above discussion, 
the disk structure around WL~17 may be interpreted as the disk with $\alpha_{\rm MRI} \lesssim 10^{-3}$, $\Cw\sim10^{-4}$, and St$\sim 0.1$.
In the ring at $t\sim 0.6$~Myr, Stokes number St=0.1 corresponds to the dust size of a few mm, which is consistent with the fact that WL~17 is observed in the mm-wavelength.
These parameters correspond to the viscous timescale $t_{\rm vis}\gtrsim 10^5$ yr, the wind timescale $t_{\rm wind }\sim 10^5$ yr, and the radial drift timescale is $t_{\rm drift } \sim 10^4$ yr.
According to Figure \ref{fig:bbl_ratio}, the dust-to-gas mass ratio is larger than 0.1 if the structure is formed by the disk wind.

\subsection{Evolution of the dust in the disk }
In this work, we treat the Stokes number (or dust size) as a parameter.
In principle, however, it should be solved simultaneously with the disk evolution.
As discussed in Section \ref{a_dust}, it is difficult to form the ring structure with an inner hole in the dust disk by using the models with a constant dust radius.
Thus, dust growth might be important to explain the ring structure like WL~17.
As zeroth order approximation, we take into account the growth of the dust by using the models with constant Stokes number according to the one-dimensional calculation of the dust growth \cite[]{2012ApJ...752..106O}.
Since the growth timescale of the dust is about $3\times 10^3$ yr at 10 au \cite[cf.][]{1996ApJ...460..832T}, we consider that the assumption that the dust grows to St=0.1 within the disk evolution timescale $\sim$ 0.1~Myr is not too unrealistic.
However, there might be some problem in this treatment.
The most serious problem will be that the constant Stokes number corresponds to the decrease of the dust radius when the surface density of gas decreases due to the wind mass loss.
The constant Stokes number assumption may be valid if the destruction of the dust due to the collision or supplying the small dust from the outer radius by the drift of the dust makes the dust radius decrease.
If the dust radius does not change when the gas surface density decreases, Stokes number of the dust increases.
This makes the dust concentration on the pressure maximum faster than the results presented in this paper.
Thus, the condition of the ring structure formation may not be too affected, or it may be possible to form ring structures with smaller initial Stokes number.

\subsection{Effects of magnetic field on disk formation and evolution}

In our model, we assume the conservation of the angular momentum in the infalling envelope for simplicity. 
When the cloud core is magnetized, however, the angular momentum is transferred outward in the envelope. 
This is called ``magnetic braking''.
The initial configuration of a cloud core and non-ideal MHD processes both affect the angular momentum transfer in the infalling envelope \cite[cf.][]{2014prpl.conf..173L,2014MNRAS.438.2278M,2015ApJ...810L..26T}, while we still do not have quantitative models for these effects.
Since the angular momentum transfer in the envelope affects the disk radius, we need further investigation of modeling of the gravitational collapse of the cloud core taking into account the magnetic braking.

The strength of the turbulence and the efficiency of the wind mass loss depend on the magnetic field in the disk.
Thus, in principle, these parameters cannot be chosen independently but related to each other.
The assumption that $\alpha_{\rm MRI}$ and $\Cw$ are constant corresponds to the smooth distribution of the magnetic field in the disk.
However, it depends on the radius in reality \cite[cf.][]{2014ApJ...784..121S,2014ApJ...796...31B,2015A&A...574A..68F}.
Although the $\alpha_{\rm MRI}$ and $\Cw$ depend on the radius of the disk, the condition for the ring structure formation based on the comparison of the timescales obtained in this work will be available. 
According to \cite{2010ApJ...718.1289S}, $\alpha_{\rm MRI}$ is about an order of magnitude larger than $\Cw$ (Figure 8 in \cite{2010ApJ...718.1289S}, see also \cite{2011ApJ...742...65O}).
We have observed the inner hole formation when $\alpha_{\rm MRI}\leq3\times10^{-4}$ and $\Cw=10^{-4}$.  
These values are roughly consistent with or the wind mass loss is slightly (a factor of a few) more effective than those obtained in the local simulations.  

The non-ideal MHD effects are also important for the disk structure formed by the disk wind. 
In \cite{2010ApJ...718.1289S}, only the Ohmic resistivity is taken into account. 
As a result, the layered structure consisting of the MRI-inactive midplane and the MRI-active surface is realized.
On the other hand, the MRI at the surface layer is suppressed by the ambipolar diffusion \cite[]{2013ApJ...769...76B,2013ApJ...772...96B,2015ApJ...801...84G} and the accretion flow is laminar in the inner region of the disk.
In such a case, the disk accretes due to the wind torque.

Moreover, the Hall effects also changes the angular momentum transfer and wind mass loss rate \cite[]{2014ApJ...791..137B,2015ApJ...798...84B,2014A&A...566A..56L,2015MNRAS.454.1117S}.
In the inner disks ($\lesssim 10-30$ au), the disk evolution depends on the orientation of the magnetic field to the rotation axis of the disk, ${\bf \Omega \cdot B}$.
If ${\bf \Omega \cdot B}>0$, the azimuthal magnetic field is amplified leading to the efficient angular momentum transfer and wind mass loss.
On the other hand, if ${\bf \Omega \cdot B}<0$, the horizontal magnetic field is reduced and the Maxwell stress in the disk and wind mass loss rate decreases.
Thus, the ring structure formation in a young disk will be difficult for the disk with ${\bf \Omega \cdot B}<0$ because the wind mass loss timescale may be longer than the age of the disk.
In the outer disk ($\gtrsim 30$ au), the surface layer of the disk is ionized with FUV irradiation and MRI turbulence is sustained there.
As a result, efficient mass accretion is expected at outer radii.
If the gas from the outer disk piles up at the boundary of the inner region, the additional ring structure might be formed.

Although the ring structure formation is affected by the complicated non-ideal MHD effects, the effects have not been fully understood quantitatively.
Further investigations of global, 3D simulations on magnetorotational instability are necessary to construct more realistic models of early disk evolution.

\section{Conclusion}
\label{conclusion}
In this work, we investigate the ring structure formation in young disks by the wind mass loss caused by the MRI turbulence.
To calculate the ring formation in young disks, we use a one-dimensional disk model that treats the formation and evolution of disks in a single framework.
In this model, the strength of the turbulence, the mass loss rate by the disk wind, and dust size (Stokes number) are treated as parameters, and the dependence of the disk evolution on these parameters are investigated. 
Main results obtained in this work are summarized as follows:
\begin{itemize}
 \item 
We find five types of disk structures as a result of disk formation and evolution model including the effects of wind mass loss: the ring structure, the dust gap disk, the filled disk, the dust poor disk, and the normal filled disk are obtained from our model calculations.
 \item The disk evolution is characterized by the timescales of the viscous diffusion $t_{\rm vis}$, the wind mass loss $t_{\rm wind}$, and the radial drift of the dust $t_{\rm drift}$.
The formation of various disk structures can be understood by comparing these timescales.
 \item 
The relation between the timescales and the disk structure is summarized in Figure \ref{fig:type_table}.
When $t_{\rm wind}<t_{\rm vis}$, the inner hole structure is formed in the gas disk.
In this case, the dust can concentrate at the pressure maximum of the gas disk and the ring structure is formed when $t_{\rm drift} < t_{\rm wind}< t_{\rm vis}$. 
In the case $t_{\rm wind}< t_{\rm vis} < t_{\rm drift}$ or $t_{\rm wind}< t_{\rm drift} < t_{\rm vis}$ in the entire disk, the gas inner hole expands faster than the radial drift of the dust. 
As a result, the dust disk remains in the inner hole of the gas disk and the filled dust disk is formed.
If the disk satisfies $t_{\rm wind}< t_{\rm vis} < t_{\rm drift}$ or $t_{\rm wind}< t_{\rm drift} < t_{\rm vis} $ only in the inner region and the outer region satisfies $t_{\rm drift} < t_{\rm wind}< t_{\rm vis}$, the dust gap disk is formed.
When $t_{\rm vis}<t_{\rm wind}$, the inner hole structure of the gas disk is not formed.
In this case, the dust ring structure is not formed either.
If $t_{\rm drift}<t_{\rm vis}<t_{\rm wind}$, the dust in the disk drifts inward faster than the gas disk evolution.
As the result, the dust poor disk is formed.
On the other hand, the dust-to-gas mass ratio is almost same as the initial value (0.01) if $t_{\rm vis}<t_{\rm wind} < t_{\rm drift}$ or $t_{\rm vis}< t_{\rm drift}< t_{\rm wind}$.
This case is categorized as the normal filled disk.
\item When the ring structure is formed, the ring radius and the dust-to-gas mass ratio at the radius of the dust surface density maximum increase with time.
When the dust-to-gas mass ratio becomes larger than about unity, the back reaction from the dust to the gas becomes efficient and the ring radius does not increase after that.
\item To explain the ring structure observed around WL~17, $\alpha_{\rm MRI} \lesssim 10^{-3}$, $\Cw \sim 10^{-4}$ and St$\sim 0.1$ are suitable based on our model calculations.
The resultant structures obtained by using these parameters are consistent with the ring radius, the dust surface density of the ring, formation of the inner hole in the dust disk, and mass loss rate estimated from observations.
Our model calculations suggest that the dust-to-gas mass ratio at the ring is larger than 0.1 and the outer radius of the gas disk is larger than that of the dust ring.
\end{itemize}

In reality, the dust radius, strength of the turbulence, and the mass loss rate due to the wind are given by the result of the dust growth and the evolution of the magnetic field in the disk.
The calculation of these values with the disk evolution is required to obtain more realistic structure of the disk, which will be the subject of our future work.

We thank Takeru K. Suzuki for fruitful discussions and his valuable comments.
This work was supported by NAOJ ALMA Scientific Research Grant Numbers 2016-02A.
TM is supported by JSPS KAKENHI Grant Nos. 17H01103, 15H02074, and 26800106.

\appendix
\section{Derivation of the velocity of the gas and dust in the disk}
The equations for the gas and the dust in the disks are given as follows:
\begin{equation}
 \deldel{\Sigma}{t} = -\frac{1}{r}\deldel{}{r}(r \Sigma u_r) + {\dot \Sigma}_{\rm inf} -{\dot \Sigma}_{\rm wind},
\label{eq:eoc_g}
\end{equation}
\begin{equation}
 \Sigma \left(\deldel{u_r}{t}+u_r\deldel{u_r}{r}-\frac{u_\phi^2}{r}\right)
=-\deldel{P}{r}-\Sigma\frac{GM_r}{r^2}+\frac{v_r-u_r}{\ts}\Sigma_{\rm d},
\label{eq:eom_r_g}
\end{equation}
\begin{equation}
 \Sigma \left(\deldel{u_\phi}{t}+u_r\deldel{u_\phi}{r}
+\frac{u_ru_\phi}{r}\right)=
\frac{1}{r^2}\deldel{}{r}\left(\Sigma\nu r^3\deldel{\Omega}{r}\right)+\frac{v_\phi-u_\phi}{\ts}\Sigma_{\rm d},
\label{eq:eom_phi_g}
\end{equation}
\begin{equation}
 \deldel{\sd}{t} = -\frac{1}{r}\deldel{}{r}(r \sd v_r) + \epsilon{\dot \Sigma}_{\rm inf},
\label{eq:eoc_d}
\end{equation}
\begin{equation}
 \sd \left(\deldel{v_r}{t}+v_r\deldel{v_r}{r}-\frac{v_\phi^2}{r}\right)
=-\Sigma\frac{GM_r}{r^2}+\frac{u_r-v_r}{\ts}\Sigma_{\rm d},
\label{eq:eom_r_d}
\end{equation}
\begin{equation}
 \sd \left(\deldel{v_\phi}{t}+v_r\deldel{v_\phi}{r}
+\frac{v_rv_\phi}{r}\right)=
\frac{u_\phi-v_\phi}{\ts}\Sigma_{\rm d},
\label{eq:eom_phi_d}
\end{equation}
where $\Sigma$, $u_r$, $u_\phi$ are the surface density, the radial velocity, and the azimuthal velocity of the gas, $\sd$ , $v_r$, $v_\phi$ are those of the dust, ${\dot \Sigma}_{\rm wind}$ is the mass loss rate due to the disk wind, $P$ is the vertically integrated pressure, $M_r$ is the enclosed mass of the gas within the radius $r$, $\ts$ is the stopping time of the dust, $\nu$ is the coefficient of the kinematic viscosity, and $\epsilon$ is the dust-to-gas mass ratio in the infalling envelope.

We assume that the effect of the pressure gradient force, the frictional force between the gas and the dust, and the viscosity is small compared with the centrifugal force and the gravitational force. 
If we neglect these forces, we obtain $u_r=v_r=0$ and $u_\phi=v_\phi=\sqrt{GM_r/r}$. 
We calculate the deviation of the velocities from these values in the first order. 
Here we treat that $\deldel{}{t}$ is the same order as $v_r\deldel{}{r}$.
We define $\delta u_\phi$ and $\delta v_\phi$ as the deviations of the azimuthal velocities.
From Equation (\ref{eq:eom_r_g}) and (\ref{eq:eom_r_d}), we obtain
\begin{equation}
 -\Sigma\frac{2u_\phi\delta u_\phi}{r} = -\deldel{P}{r} +\frac{v_r-u_r}{\ts}\sd,
\label{eq:eom_r_g_1}
\end{equation}
\begin{equation}
 -\sd\frac{2v_\phi\delta v_\phi}{r} = \frac{u_r-v_r}{\ts}\sd.
\label{eq:eom_r_d_1}
\end{equation}
Equation (\ref{eq:eom_r_g_1}) gives 
\begin{equation}
 \delta u_\phi = -\frac{1}{2}\frac{\cs^2}{u_\phi^2}
\left|\frac{r}{\cs^2\Sigma}\deldel{P}{r}\right|u_\phi
-\frac{1}{2}\frac{\sd}{\Sigma}\frac{v_r-u_r}{\St},
\end{equation}
where $\St=\ts \sqrt{GM_r/r^3}$.
To obtain the radial drift of the dust in the disk midplane, we evaluate the first term by using $\eta$ defined in the midplane, where $\eta=-1/2(\cs/u_\phi)^2d\ln p/d\ln r$ and $p$ is the pressure in the midplane.
Equation (\ref{eq:eom_r_d_1}) gives 
\begin{equation}
 \delta v_\phi = -\frac{1}{2}\frac{u_r-v_r}{\St}.
\end{equation}
Thus, we obtain
\begin{equation}
 \delta v_\phi-\delta u_\phi =\eta u_\phi
+\frac{1}{2}\frac{v_r-u_r}{\St'},
\end{equation}
where
\begin{equation}
 \St'= \frac{\Sigma}{\Sigma +\sd}\St.
\end{equation}
From Equations (\ref{eq:eom_phi_g}) and (\ref{eq:eom_phi_d}),
\begin{equation}
 r\Sigma\deldel{}{t}ru_\phi+ru_r\Sigma\deldel{}{r}ru_\phi = r\Sigma N 
+\frac{\delta v_\phi-\delta u_\phi}{\ts}r^2\sd,
\label{eq:eom_g_phi_1}
\end{equation}
\begin{equation}
 r\sd\deldel{}{t}rv_\phi+rv_r\sd\deldel{}{r}rv_\phi = \frac{\delta u_\phi-\delta v_\phi}{\ts}r^2\sd,
\label{eq:eom_d_phi_1}
\end{equation}
where
\begin{equation}
 N=\frac{1}{r\Sigma}\deldel{}{r}\left(
r^3\nu\Sigma\deldel{\Omega}{r}
\right)
\end{equation}
is the specific torque and $\Omega=u_\phi/r=v_\phi/r$.
We define the specific angular momentum $j=ru_\phi=rv_\phi$, where
\begin{equation}
 \deldel{j}{t} = \frac{1}{2}\frac{j}{M_r}
\deldel{M_r}{t},
\end{equation}
\begin{equation}
 \deldel{j}{r} = \frac{j}{2r}
\left(1+\frac{2\pi r^2 \Sigma}{M_r}\right).
\end{equation}

The time derivative of the enclosed mass $M_r$ is given by the integration of Equation (\ref{eq:eoc_g}),
\begin{eqnarray}
 \deldel{M_r}{t} &=& \int^r_02\pi r 
\left({\dot \Sigma}_{\rm inf}-{\dot \Sigma}_{\rm wind} 
-\frac{1}{r}\deldel{}{r}ru_r\Sigma\right)dr \nonumber
\\
&=&-2\pi r \Sigma u_r + {\dot M}_{r,{\rm tot}},
\end{eqnarray}
where
\begin{equation}
 {\dot M}_{r,{\rm tot}}= 
\int^r_02\pi r 
({\dot \Sigma}_{\rm inf}-{\dot \Sigma}_{\rm wind})dr 
\end{equation}
From Equation (\ref{eq:eom_g_phi_1}), we obtain
\begin{equation}
r\Sigma\frac{j}{2r}\left(u_r+\frac{r\mdrtot}{M_r}\right) = r\Sigma N 
+\frac{r^2\sd}{\ts}\left(\eta r\Omega+ \frac{1}{2}
\frac{v_r-u_r}{\St'}\right).
\label{eq:eom_r_g_2}
\end{equation}
From Equation (\ref{eq:eom_d_phi_1}), we obtain
\begin{eqnarray}
r\sd\frac{j}{2r}\left(v_r+\frac{r\mdrtot}{M_r}\right)
-r\sd \frac{j}{2M_r}(2\pi r \Sigma) (u_r-v_r)
\nonumber \\ 
= -\frac{r^2\sd}{\ts}\left(\eta r\Omega+ \frac{1}{2}
\frac{v_r-u_r}{\St'}\right)
\label{eq:eom_r_d_2}
\end{eqnarray}
The velocity $-r\mdrtot/M_r$ represents the motion caused by change of the mass and the angular momentum distribution due to the mass infall or wind mass loss.
We define the velocity except this effect: 
\begin{equation}
{\tilde u}_r=u_r+\frac{r\mdrtot}{M_r}
\end{equation}
\begin{equation}
{\tilde v}_r=v_r+\frac{r\mdrtot}{M_r}
\end{equation}
The sum of the equations (\ref{eq:eom_r_g_2}) and 
(\ref{eq:eom_r_d_2}) gives 
\begin{equation}
\tilvr-\tilur
= \frac{2r}{Aj}\frac{\Sigma}{\sd} N
-\frac{\Sigma+\sd}{\sd}\frac{\tilur}{A},
\label{eq:vr-ur}
\end{equation}
where
\begin{equation}
A= 1+\frac{2\pi r^2\Sigma}{M_r}.
\end{equation}
Substitution of this in Equation (\ref{eq:eom_r_g_2}) gives
\begin{equation}
 \tilur=\frac{2r}{j}N-\frac{\sd}{\Sigma+\sd}\frac{1}{A\St'^2+1}
\frac{2r}{j}N
+\frac{2\sd}{\Sigma+\sd}\frac{A\St'}{A\St'^2+1}\eta r \Omega
\end{equation}
Substituting this in Equation (\ref{eq:vr-ur}), we obtain
\begin{eqnarray}
 \tilvr & =& \frac{\Sigma}{\Sigma+\sd}\frac{1}{A\St'^2+1}
\frac{2r}{Aj}N
-\frac{\Sigma}{\Sigma+\sd}\frac{2\St'}{A\St'^2+1}\eta r \Omega \nonumber \\
&&+\left(\frac{2r}{Aj}N-\frac{\sd}{\Sigma+\sd}\frac{1}{A\St'^2+1}\frac{2r}{Aj}N  +\frac{2\sd}{\Sigma+\sd}\frac{\St'}{A\St'^2+1}\eta r \Omega \right)\frac{2\pi r^2 \Sigma}{M_r}.
\end{eqnarray}
From these equations, we obtain the radial velocity  of the gas and the dust $ u_r=\tilur - r\mdrtot/M_r$, $v_r=\tilvr - r\mdrtot/M_r$.
We calculate the evolution of the surface densities of the gas and the dust by using equations (\ref{eq:eoc_g}) and (\ref{eq:eoc_d}) with these radial velocities.
The equations derived here are the same as those obtained in \cite{2009ApJ...690..407K} and \cite{2017ApJ...844..142K} if we neglect the effect of the disk self-gravity and the input/loss of the gas and the dust.

\bibliographystyle{apj}

\end{document}